\documentclass[a4paper,oneside,final,notitlepage,onecolumn,11pt]{article}

\usepackage{ifpdf,epsfig,array,amsmath,amssymb,psfrag}
\usepackage{ulem} 
\ifpdf
\usepackage[colorlinks=true,
filecolor=blue,linkcolor=blue,citecolor=blue,urlcolor=blue
]{hyperref}
\usepackage[usenames,dvipsnames]{color}
\else
\usepackage[usenames,dvips]{color}
\fi

\DeclareFontFamily{OT1}{rsfs}{} \DeclareFontShape{OT1}{rsfs}{m}{n}{
<-7> rsfs5 <7-10> rsfs7 <10-> rsfs10}{}
\DeclareMathAlphabet{\mycal}{OT1}{rsfs}{m}{n}

\def\scri{{\mycal I}}
\def\scrip{\scri^{+}}%

\DeclareFontFamily{OT1}{rsfs}{} \DeclareFontShape{OT1}{rsfs}{m}{n}{
<-7> rsfs5 <7-10> rsfs7 <10-> rsfs10}{}
\DeclareMathAlphabet{\mathscr}{OT1}{rsfs}{m}{n}

\def\eeepsilon{{\epsilon}}

\newcommand\sgn{\mathop{\rm sgn}\nolimits}
\newcommand\ds{\mathrm{d}s}
\newcommand\dt{\mathrm{d}t}
\newcommand\dtau{\mathrm{d}\tau}
\newcommand\dT{\mathrm{d}T}
\newcommand\dR{\mathrm{d}R}
\newcommand\dr{\mathrm{d}r}

\newcommand\drho{\mathrm{d}\rho}
\newcommand\dtheta{\mathrm{d}\vartheta}
\newcommand\dvarphi{\mathrm{d}\varphi}

\newcommand\gotR{\alpha}

\begin{document}

\title{Gravitational collapse and topology change in 
  spherically symmetric dynamical systems}


\author{\small 
P{\'e}ter Csizmadia\thanks{email: cspeter@rmki.kfki.hu}
\ and 
Istv\'{a}n R\'{a}cz\thanks{email: iracz@rmki.kfki.hu}
\\ 
\small RMKI\\ 
\small  H-1121 Budapest, Konkoly Thege Mikl\'os \'ut 29-33.\\ 
\small Hungary
}
\maketitle

\begin{abstract}
A new numerical framework, based on the use of a simple first order strongly
hyperbolic evolution equations, is introduced and tested in case of
$4$-dimensional spherically symmetric gravitating systems. The analytic setup
is chosen such that our numerical method is capable to follow the time
evolution even after the appearance of trapped surfaces, more importantly,
until the true physical singularities are reached. Using this framework, the
gravitational collapse of various gravity-matter systems are investigated,
with distinguished attention to the evolution in trapped regions.
It is justified that in advance to the formation of these curvature
singularities, trapped regions develop in all cases, thereby supporting the
validity of the weak cosmic censor hypothesis of Penrose.  Various upper
bounds on the rate of blow-up of the Ricci and Kretschmann scalars and the
Misner-Sharp mass are provided.  In spite of the unboundedness of the Ricci
scalar, the Einstein-Hilbert action was found to remain finite in all the
investigated cases. In addition, important conceptual issues related to the
phenomenon of topology changes are also discussed.

\end{abstract}

\vfill\eject

\section{Introduction}

The diffeomorphism invariance of Einstein's theory of gravity is in an
intimate relation with the fact that
there is a significant redundancy in the representation of the true
physical degrees of freedom. It can therefore be a great challenge to carry
out a faithful investigation of dynamical processes even in case of
spherically symmetric spacetimes, in spite of the simplifications offered by
the symmetries. Correspondingly, the selection of the most appropriate
variables, done by applying a suitable gauge fixing, i.e., the most effective
framework to carry out the study of a given dynamical system, is considered to
be a sort of art.  

In the numerical investigations of spherically symmetric
dynamical systems, the method of Choptuik---that had been applied first by him
in \cite{ch1} while exploring the critical phenomenon in the gravitational
collapse of various gravity-matter systems---turned to be the most successful
in the sense that it is still widely used.
However, Choptuik's choice has both advantages and disadvantages.
Perhaps the most important advantage is that one has to solve only two first
order partial differential equations (PDEs) for the basic metric variables
\footnote{In this particular case, the basic metric variables are the smooth
functions $A$ and $B$ in terms of which the spacetime metric can be given as
\begin{equation}
\ds^2\,\ =\,\ A\,\dt^2 - B\,\dr^2 - r^2\left(\dtheta^2 +
\sin^2{\hskip-.07cm}\vartheta\,\dvarphi^2\right)\,.
\label{the_metric_Ch.eq}
\end{equation}
}, which, along with the
matter field equations, determine the full evolution of the associated gravity
matter system. On the other hand, the following objections may also be
raised. First of all, the radial coordinate $r$ is chosen so that the area
$\mathcal{A}$ of the $SO(3)$-invariant $2$-spheres are given as
$\mathcal{A}=4\pi r^2$. As it is well-known, coordinate systems of this type
are not suitable to follow evolution in regions where ``trapped surfaces'' are
formed because a coordinate singularity also develops (for a short discussion
see, e.g., the last paragraph of \cite{Racz}). 
 
\medskip

Another objection is that one of the first order PDEs is hyperbolic while the
other is elliptic. This means that one of the metric equations---which is a
constraint equation---has to be integrated on succeeding time level surfaces
repeatedly. This process slows down time integration and makes it hard to
apply the powerful tool of adaptive mesh refinement (AMR) which ensures high
numerical accuracy in strongly dynamical processes \cite{BO,csp,csp2}. There
are two side-remarks in order. First of all, although it is possible to
implement a variant of AMR for the numerical integration of mixed hyperbolic
and  elliptic equations \cite{PretoriusChoptuik}, the precision is partly lost
because the applicability of AMR necessitates the extrapolation of some
variables in time, something which is better to be avoided in a time
integration process.  Thereby, the use of a fully hyperbolic system is
preferable in case of numerical integration of the field equations based on a
finite difference schema. Secondly, it was shown by one of the present authors
in \cite{Racz} that by making use of the Kodama  vector field as the time
evolutions vector field, the mixed elliptic-hyperbolic system can be replaced
by a fully hyperbolic one. However, we have found the inevitable formation of
a coordinate singularity in numerical simulations, the appearance of which was
anticipated in the discussion below Equation (4.15) of \cite{Racz}. The above
findings motivate the search for a better analytic set-up with a more
appropriate choice of the basic variables.
 
\medskip
      
Before proceeding and presenting our proposal for such a choice we would
like to mention that several attempts have also been made to  use reduced
versions of the  ADM and BSSN \cite{BSSN2,BSSN1} formalisms in
spherical symmetry, see \cite{B1,SF1,SF2}.
These models have been yielded by the reduction of a more complicated
evolutionary system, therefore they do not optimally fit the spherically
symmetric setup. Either they apply coordinates that are not suitable to cover
regions with trapped surfaces, as it happens, e.g., in \cite{SF1,SF2}, or they
are simply too complicated---see, e.g., the basic set of field equations
(9a)--(9f) and (10a)--(10c) in \cite{B1}. They may also suffer from numerical
instabilities at the origin because negative powers of $r$ appear in the
evolutionary equations.
 
\medskip

Thereby, it is of considerable interest to single out a simple and general
enough framework within which time evolution can be investigated on the
largest possible part of the physical spacetime up to the appearance of true
geometrical singularities. To match this requirement, we shall start by
choosing a fully hyperbolic evolutionary system that is automatically
applicable to describe the evolution in trapped region(s). This choice should
be such that the equations are free from the numerical instabilities that used
to appear in the origin in spherically symmetric spacetimes.
 
\medskip

An analytic framework fitting the above outlined expectations may be chosen
as follows. Based on the results of earlier investigations in spherically
symmetric (see, e.g., \cite{kodama,christ1,dafermos1}) and also in generic
(see, e.g., \cite{sanchez}) dynamical spacetimes, the metric of the four
dimensional spherically symmetric spacetime $(M,g_{ab})$ will be assumed to
possess the form
\begin{equation}
\ds^2\,\ =\,\ \gotR\,\beta^2\,\dtau^2 - \gotR\,\drho^2
	- r^2\left(\dtheta^2 +
        \sin^2{\hskip-.07cm}\vartheta\,\dvarphi^2\right)\,, 
\label{the_metric.eq}
\end{equation}
where the coordinates $\tau$ and $\rho$ label the points of the
two-dimensional timelike surfaces transverse to the transitivity surfaces of
the rotation group, and $\gotR$, $\beta$, $r$ are smooth functions of ($\tau$,
$\rho$).
Note that $\rho$ generally differs from the area-radial coordinate. This
condition, as we have already mentioned, is necessary to extend the domain of
time evolution to include trapped surfaces---when they exist.

\medskip

\begin{figure}[ht]
 \centerline{
  \epsfxsize=13cm 
\epsfbox{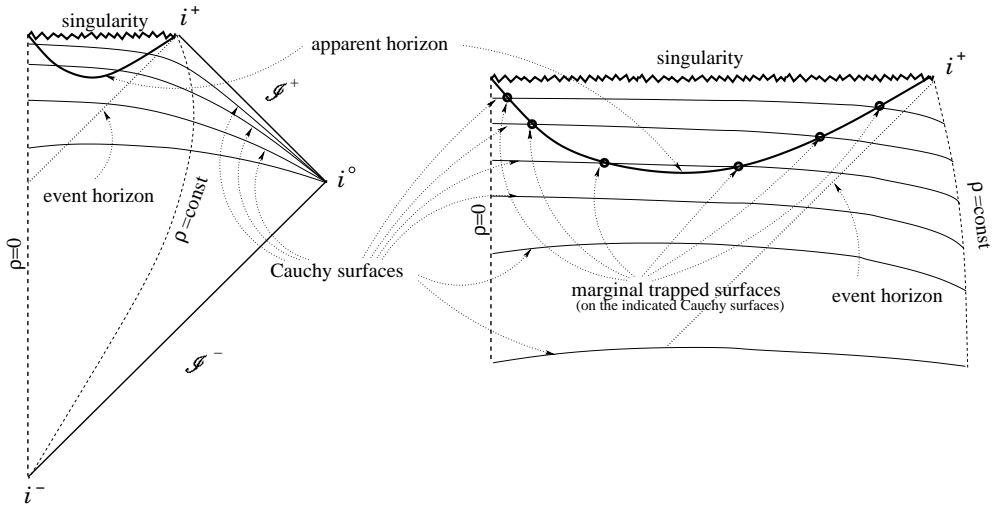}
 }
\caption{\footnotesize \label{coll} A typical spacetime diagram representing
  the gravitational collapse of a simple spherically symmetric gravity-matter
  system. In our numerical investigations, we shall focus on the
  characterisation of a gravitational collapse by monitoring
  the intersections of the
  Cauchy surfaces and the apparent horizon---indicated by circles.
  }
\end{figure}

By making use of this geometrical framework, gravitational collapse have been
investigated in some simple gravity-matter systems. In the simplest possible
case of asymptotically flat configurations, the associated time evolution of
the system is qualitatively expected
\cite{christ0,christ1,christ2,dafermos1,dafermos2} to be as indicated on
Fig.\,\ref{coll} where the event horizon\footnote{
The event horizon is the boundary of the causal past of future null infinity,
$\mycal{H}=\partial J^-[\scrip]$.},
and the apparent horizon are also shown.
The latter is foliated by marginally trapped surfaces and it is represented by
a curve connecting the two ends of the zigzag line depicting the singularity.
A portion of matter either falls into the singularity or reaches future
timelike or null infinity, $i^+$ or $\scrip$.
The asymptotic structure of the spacetime is expected to approach that of the
Schwarzschild solution as we get closer to $i^+$ along the null generators of
$\scrip$ \cite{christ1}.
If the collapsing matter has no considerable radiative degrees of freedom, the
mass of the developing black hole is expected to be close to the total mass on
the initial data surface. Note that spherical symmetry makes it possible to
define mass and energy inside an invariant metric sphere (see Section\,\ref{trap}).
If the mass inside the marginally outer trapped surface is found to approach the
total mass while moving outwards along the apparent horizon, then we have a
strong indication that our spacetime grid covers the truly dynamical part of the
collapse. Moreover, as it will be demonstrated in Section\,\ref{dyn},
the lapse function $\beta$ may always be chosen such that the
Cauchy surfaces can get arbitrarily close to the singularity.
 
\medskip

We were also interested in investigating the time evolution of more exotic
initial data specifications. Likewise in the standard
Friedman-Robertson-Walker cosmological models---these are known to be
spherically symmetric around any of their spacetime events--- there is a
freedom in choosing the topology of the initial data surfaces.

\begin{figure}[ht]
 \centerline{
  \epsfxsize=4.3cm 
\epsfbox{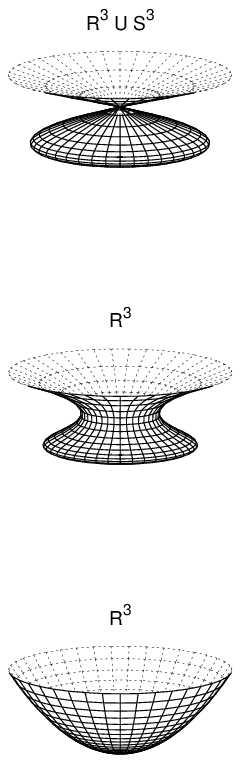}
  \epsfxsize=4.3cm 
\epsfbox{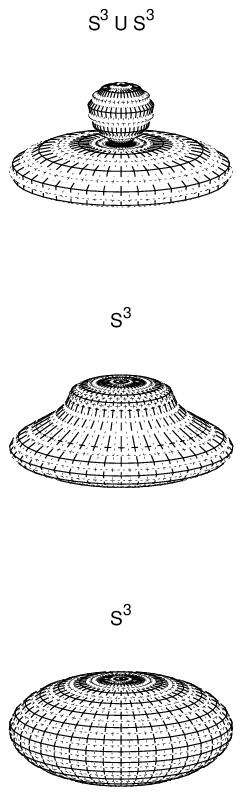}
  \epsfxsize=4.3cm 
\epsfbox{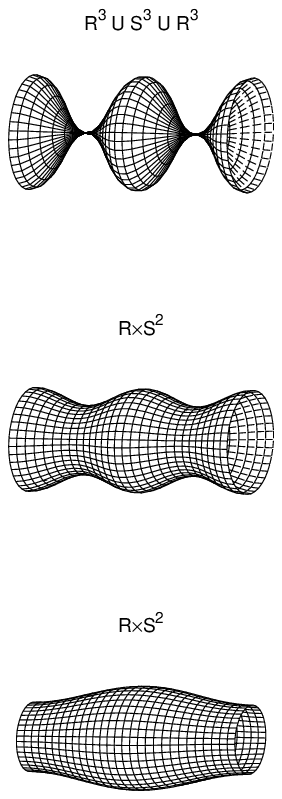}
  \epsfxsize=4.3cm 
\epsfbox{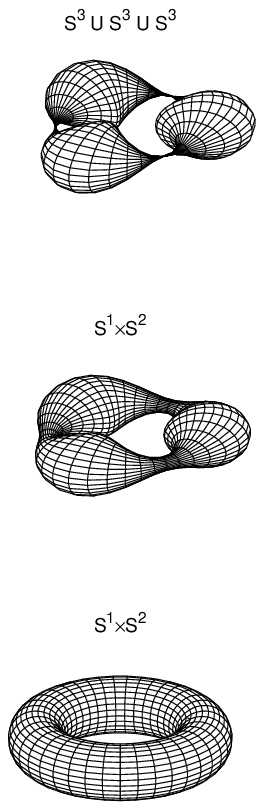}
 }
\caption{\footnotesize \label{top} Topology changes of the time level surfaces
are illustrated by the vertical figure sequences, with the suppression of one
space dimension. Time progresses from bottom to top. Transitivity surfaces of
the rotation group are represented by horizontal circles in the first two
sequences (starting from $\mathbb{R}^3$ and $\mathbb{S}^3$) and by vertical
circles in the last two sequences ($\mathbb{R}\times\mathbb{S}^2$ and
$\mathbb{S}^1\times\mathbb{S}^2$).}
\end{figure}

The base manifold $M$ of the investigated spacetimes coincides with the future
Cauchy development of some three-dimensional achronal hypersurface $\Sigma$.
Thereby, $M=D^+[\Sigma]$ and it possesses the product space structure
$\Sigma\times \mathbb{R}^+$ \cite{geroch2}.
Since the spacetime is spherically symmetric, the $\tau=const$ time level
surfaces---these are diffeomorphic to  $\Sigma$---can also be foliated by the
transitivity surfaces of the rotation group. In virtue of the particular form
of the applied line element (\ref{the_metric.eq}) the metric induced on the
transitivity surfaces of the rotation group is given as
$ds^2|_\mycal{S}=r^2\left(\dtheta^2 +
\sin^2\hskip-.07cm\vartheta\,\dvarphi^2\right)$. Thereby, the vanishing of
$r$, which is, in fact, the area-radius function, is directly related to the
existence of an origin. To see how many origins we might have, let us
consider some simple choices for the topology of the initial data surface
$\Sigma$.
Whenever $\Sigma$ is a connected geodesically complete spacelike hypersurface
possessing a trivial product bundle structure, its topology is either
$\mathbb{R}^3=[0,\infty)\times\mathbb{S}^2$,
$\mathbb{S}^3=[0,\pi]\times\mathbb{S}^2$, $\mathbb{R}\times \mathbb{S}^2$ or
$\mathbb{S}^1\times \mathbb{S}^2$ (see as an illustration the bottom line of
Fig.\,\ref{top}).  Accordingly, there may be one origin, two origins or no
origin at all on our initial hypersurface.
During time evolution, the geometrical properties of the time level surfaces
may change such that new origins are produced, possibly indicating a
change in the topology, as illustrated on Fig.\,\ref{top}.

\medskip 

At this point it is important to clear up the some of the related notions and
potential misconceptions. While in the above sentences the terminology of
topology change have been used, it should be kept in mind that it is the
topology of the limit of time level surfaces---comprising the future
boundary of the pertinent ``domain of dependence''---what may differ from
that of the time level surfaces. A careful approach is needed in applying
notions such as {\it domain of dependence, Cauchy development} and {\it Cauchy
surface} which were introduced in \cite{geroch2} by Geroch. We would like to
emphasise first that there is a significant distinction between the notions
``initial data surface'' and ``Cauchy surface''.  We may chose an arbitrary
achronal hypersurface $\Sigma$ as an initial data surface. The initial data
together with the field equations can be used to determine the solution
everywhere in the domain of dependence $D[\Sigma]$. As opposed to this, the
use of the notion of ``Cauchy surface'' tacitly requires additional knowledge
about the global properties of the underlying spacetime. In particular, see
Theorem 11 of \cite{geroch2}, a spacetime is known to be globally hyperbolic
if and only if it possesses a Cauchy surface. 

The confusion might arise---and, in fact, apparently it does arise in many
circumstances---in consequence of the fact that the domain of dependence
$D[\Sigma]$ of any initial data surface $\Sigma$, is a globally hyperbolic
spacetime itself on its own right, i.e., $\Sigma$ is a Cauchy surface for
$D[\Sigma]$. In many cases it is obvious to recognise that there exists a
globally hyperbolic global                              extension\footnote{%
  For the precise notion of `local' and `global' extensions of spacetimes see,
  e.g., \cite{racz1,racz2}. } $(M',g'_{ab})$ of $(D[\Sigma],g_{ab})$ such that
the boundary of $D[\Sigma]$ is not empty in $M'$, while the Cauchy horizon,
$H[\Sigma]=H^+[\Sigma] \cup H^-[\Sigma]$ of $D[\Sigma]$ is obviously empty
with respect to $D[\Sigma]$. However, a generic method that could be applied
in all the possible cases does not exist.  Although the concept of
``maximal Cauchy development'' was introduced in \cite{geroch0}, it is
based on Zorn's lemma, which makes its use difficult in practice.

\begin{figure}[ht]
 \centerline{
  \epsfxsize=7cm 
\epsfbox{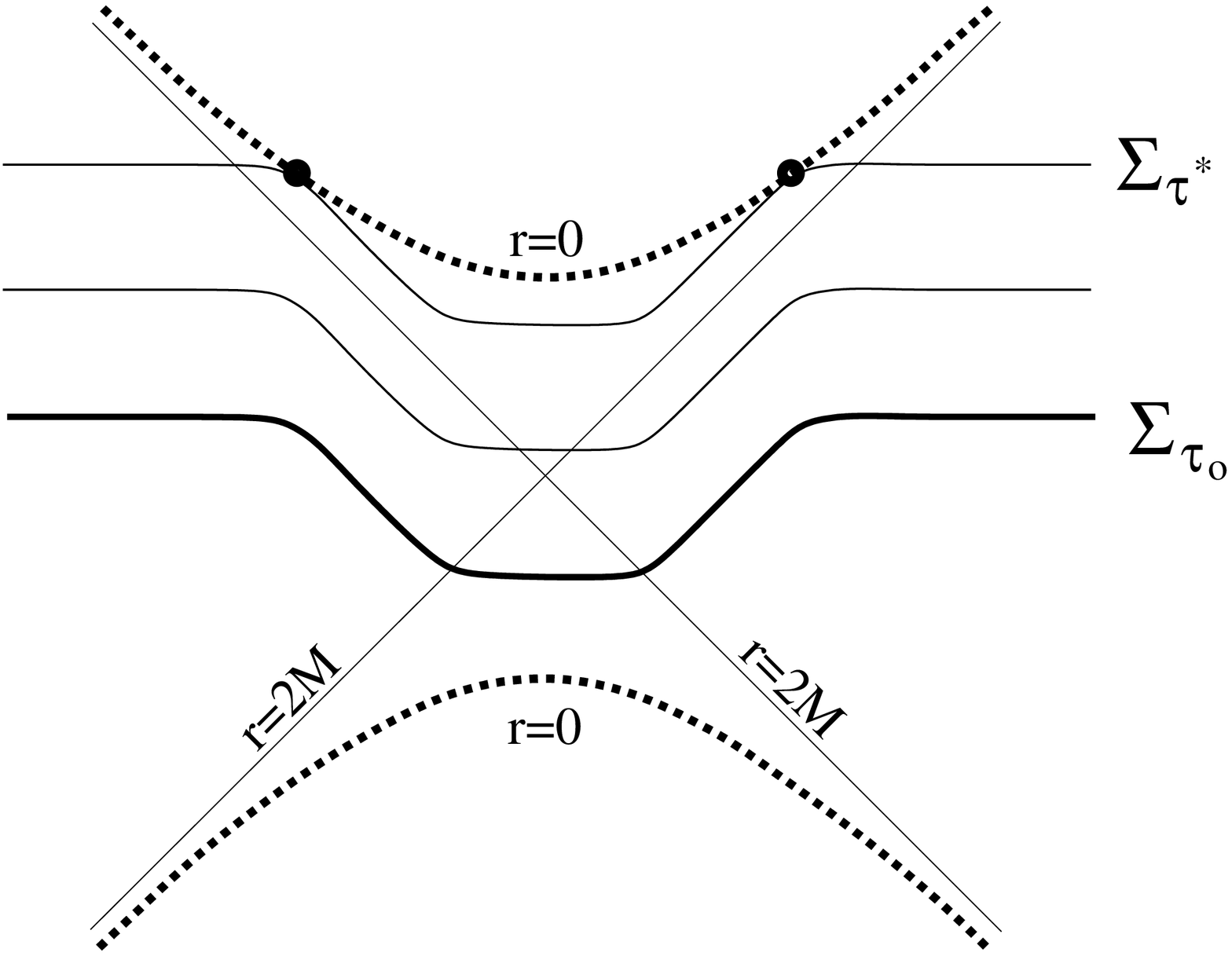}
\ \ \ \epsfbox{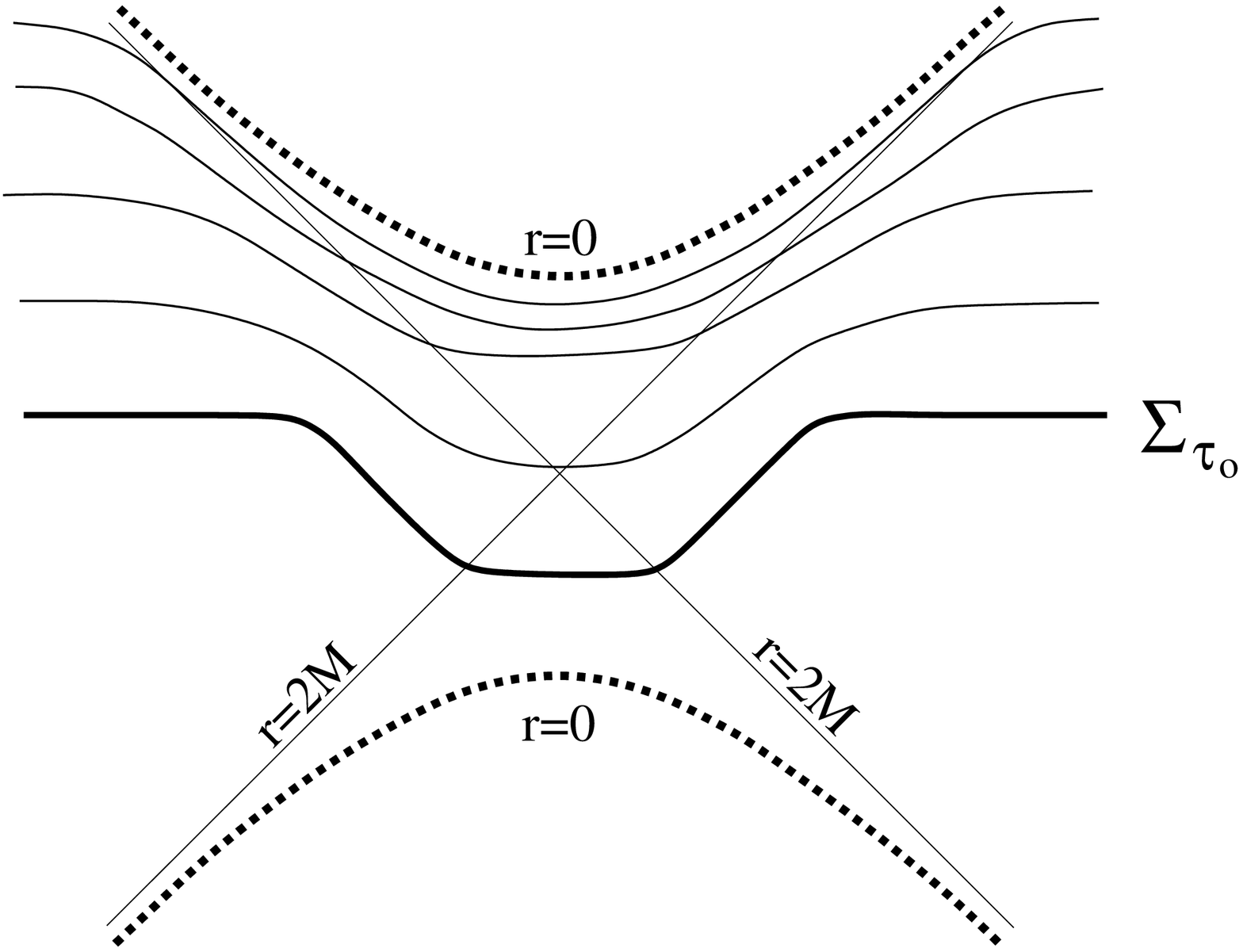}
 }
\caption{\footnotesize \label{sch} Time slicings with and without
topology change. On the left, new origins appear (full circles)
during the evolution.  The topology of $\Sigma_{\tau^*}$ differs from that of
the $\Sigma_\tau$ hypersurfaces. These time level
surfaces cover only a part of the future domain of dependence
$D^+[\Sigma_{\tau_0}]$.  On the right, it is indicated
that the entire of $D^+[\Sigma_{\tau_0}]$ can be covered by a properly chosen
time slicing such that no topology change happens.}
\end{figure}

As a simple example, consider the maximal analytic extension of the
Schwarzschild spacetime. In Kruskal coordinates $(T,R,\vartheta,\varphi)$ the
line element is given by
\begin{align}
\ds^2\, &=\ \gotR\,\dT^2 - \gotR\,\dR^2
	- r^2\left(\dtheta^2 +
        \sin^2{\hskip-.07cm}\vartheta\,\dvarphi^2\right)\,,
\label{ds2.KruskalSzekeres.eq}\\
\alpha\, &=\ \frac{32M^3e^{-r/2GM}}{r}\,,
\end{align}
where the Schwarzschild coordinates $r$ and
$t$ are determined by the following implicit relations (see, e.g.,
\cite{wald}):
\begin{eqnarray}
\left(1 - \frac{r}{2M}
\right)\,e^{r/2GM}&&\hskip-.5cm=\hskip.1cm T^2-R^2, \label{rfv}\\ 
\frac{t}{2M}&&\hskip-.5cm=\hskip.1cm
2\,\tanh^{-1}\left(\frac{T}{R}\right)\,. 
\end{eqnarray}
By introducing suitable new coordinates $\tau=\tau(T,R)$ and $\rho=\rho(T,R)$,
the
line element of the Schwarzschild spacetime preserves the form 
(\ref{the_metric.eq}). The time slicings---two of which are indicated on Fig.\,\ref{sch}---and the corresponding lapse functions
vary accordingly.  In the first case (left panel), two points on the $\Sigma_{\tau^*}$
hypersurface hit the $r=0$ Schwarzschild singularity.  The appearance of these
points can be interpreted as the formation of two origins, since $r>0$ in all
other points of $\Sigma_{\tau^*}$.
Correspondingly, the topology $\mathbb{R}\times \mathbb{S}^2$ of
$\Sigma_{\tau_0}$ is replaced by the disjoint union of
$\mathbb{R}^3=\mathbb{R}^+\times \mathbb{S}^2$, $\mathbb{S}^3$ and
$\mathbb{R}^3$ at $\tau^*$ (see the third column of Fig.~\ref{top}).
It is not hard to modify $\Sigma_{\tau_0}$ by inserting suitably located
further ``steps'' such that the number of $\mathbb{S}^3$ components of the
pertinent $\Sigma_{\tau^{*}}$ may take an arbitrary integer value.

Although the time level surfaces on the left panel do not cover the future
domain of dependence, $D^+[\Sigma_{\tau_0}]$, of $\Sigma_{\tau_0}$, it is
worthwhile to keep in mind that $\Sigma_{\tau_0}$ is a Cauchy surface for the
Schwarzschild spacetime. As it is indicated on the right panel of
Fig.\,\ref{sch}, the entire of $D^+[\Sigma_{\tau_0}]$ can be covered by time
level surfaces if the new coordinates $\tau$ and $\rho$, and in turn, the lapse
$\beta$ are chosen properly. There is no topology
change and the section of the Cauchy surfaces within the black hole region
uniformly 
converge to the $r=0$ singularity. In section\,\ref{dyn},
the application of this type of dynamically determined $\beta$ will be
demonstrated via various examples.

\bigskip

We would like to mention that the type of topology changes indicated by the
left panel of Fig.\,\ref{sch} showed up in many of our numerical simulations.
We have found that, in consequence of the developing inhomogeneities, certain
parts of the time level surfaces may get closer to the singularity much faster
than others. Nevertheless, as shown by the above example, the change of the
topology refers to the limit of the applied time level surfaces rather than
the underlying spacetimes. As it will be demonstrated in the succeeding
sections, the $\beta$ function can be chosen such that the time evolution is
slowed down in spacetime regions close to the singularity. This way we were
able to enlarge the domain of dependences significantly and to get arbitrarily
close to the developing spacetime singularities in our simulations. As a
consequence, we were able to investigate the rate of blowing up of the
curvature while approaching the singularity.

\medskip

The structure of this paper is the following. In Section\,\ref{intr}, with the
help of a simple-minded approach, we demonstrate the main technical issues in
solving the Einstein-scalar field equations numerically. These difficulties
are resolved in Section\,\ref{intr2}, where the Misner-Sharp mass is
introduced as an auxiliary variable to stabilise the applied numerical
representation near the origin. Here we also describe the ``trick'' which
makes our framework capable to follow the time evolution within the trapped
region.  Section\,\ref{stabor} outlines the analytic procedure that guarantees
the required stability of the time evolution in the neighbourhood of an
origin. In Section\,\ref{energybal}, a method is presented which provides the
opportunity to monitor the energy transfer processes. Trapped surfaces,
trapped regions and apparent horizons are defined in Section\,\ref{trap}.
Dynamical investigations of gravitational collapse are presented in simple and
in certain less obvious circumstances are presented in Section\,\ref{dyn}.
The results of the convergence and accuracy tests are also presented in this
section. The paper is closed by summarising our results and by providing a
short discussions concerning the interpretation and some of the immediate
consequences of our findings.

\section{A simple-minded approach}\label{intr}

To have a truly dynamical spacetime, some matter fields also have to be
specified in addition to the already introduced geometrical framework.  For
simplicity, we shall restrict our considerations to a real self-interacting
scalar field with Lagrangian 
\begin{align}
{\mycal L}\ =\ \frac{1}{2} g^{ef}\nabla_e\psi\nabla_f\psi  -  V,
\label{Lagrangian.scalar.eq}
\end{align}
where the potential possesses the form relevant to the Klein-Gordon field,
$V(\psi)={\frac{1}{2}}\mu^2\,\psi^2$ with mass parameter $\mu\geq 0$. Note,
however, that $V$ could be chosen to be any sufficiently regular function of
the $\psi$ field.

The field equations relevant for the gravity-matter system can be given as  
\begin{eqnarray}\label{einst}
&&E_{ab}=R_{ab}-\frac12 g_{ab}R-8\pi T_{ab}=0\,,\\
&&\hskip.8cm\nabla_e\nabla^e\psi + \frac{\partial V}{\partial\psi}=0,
\label{psi.eq}
\end{eqnarray}
where the energy-momentum tensor reads as 
\begin{eqnarray}
&&T_{ab}= \nabla_a\psi\nabla_b\psi - g_{ab}
	\left[\frac{1}{2}g^{ef}\nabla_e\psi\nabla_f\psi - V\right]\,.
\end{eqnarray}
It is straightforward to verify, by a direct calculation,
that our basic variables $\gotR$, $\beta$, $r$ and $\psi$ satisfy the
evolution equations 
\begin{align}
\partial_\tau\gotR_\tau
\,\ =\,\ &\beta^2\partial_\rho\gotR_\rho
	- 8\pi \gotR(\psi_\tau^2 - \beta^2\psi_\rho^2)
	+ \frac{\gotR_\tau^2 - \beta^2\gotR_\rho^2}{\gotR}
	\nonumber\\
& + 2\gotR\frac{\gotR\beta^2+ r_\tau^2 - \beta^2r_\rho^2}{r^2}
        +\ \frac{\beta_\tau \gotR_\tau + \beta^2 \beta_\rho\,\gotR_\rho}{\beta} 
	+ 2\gotR\beta\,\partial_\rho\beta_\rho\,,
\label{Omega_tautau.0.eq}
\end{align}
\begin{equation}
\partial_\tau r_\tau
\,\ =\,\ \beta^2\partial_\rho r_\rho + 8\pi\,r\,\gotR\beta^2\, V
	- \frac{\gotR\beta^2+r_\tau^2 - \beta^2 r_\rho^2}{r}
	+ \frac{\beta_\tau r_\tau}{\beta}\,,
\label{r_tautau.0.eq}
\end{equation}
\begin{equation}
\partial_\tau\psi_\tau
\ =\ \beta^2\partial_\rho\psi_\rho
	- 2\frac{\psi_\tau r_\tau - \beta^2\psi_\rho r_\rho}{r}
	+ \frac{\beta_\tau\psi_\tau + \beta^2\beta_\rho\psi_\rho}{\beta}
	- \gotR\beta^2\frac{\partial V}{\partial\psi}\,,
\label{psi_tautau.eq}
\end{equation}
where the abbreviating notation $f_\sigma=\partial_\sigma f$ is applied
(for $f$ = $\alpha$, $\beta$, $r$ and $\psi$).
Similarly, the constraint equations are
\begin{equation}
E^\tau_{\ \rho}\ =\ 0,\qquad E^\tau_{\ \tau}\ =\ 0,\label{constraints.eq}
\end{equation}
where
\begin{gather}
E^\tau_{\ \rho} \,=\, \frac{2\,\beta\, ({\partial_\tau r_{{\rho}} })
-r_{{\tau}}\,\gotR_{{\rho}}
-\gotR_{{\tau}}r_{{\rho}}}{{{\gotR}^{2}{\beta}^{2}r}}
-\frac{2\,r_{{\tau}}\,\beta_{{\rho}}}{{{\gotR}{\beta}^{3}r}}+ 8\,\pi
\,\frac{\psi_{{\rho}}\psi_{{\tau}}}{{{\gotR}{\beta}^{2}}}\,,\\ 
E^\tau_{\ \tau} =\, \frac{2\, \gotR\,{\beta}^{2}\,(\partial_\rho r_{{\rho}})
-r_{{\tau}}\gotR_{{\tau}}
\mbox{}-{\beta}^{2}r_{{\rho}}\gotR_{{\rho}}}{{{\gotR}^{2}{\beta}^{2}{r}}}
-\frac{{r_{{\tau}}}^{2}-{r_{{\rho}}}^{2}{\beta}^{2}+{\gotR}{\beta}^{2}}
{{{\gotR}{\beta}^{2}{r}^{2}}}+4\,\pi \frac{{\psi_{{\tau}}}^{2}
  +{\beta}^{2}{\psi_{{\rho}}}^{2}+2{\gotR}{\beta}^{2}V}{{{\gotR}{\beta}^{2}}}\,. 
\end{gather}
The number of independent equations is in accordance with the fact
that the evolution equation (\ref{psi_tautau.eq}) for $\psi$ is simply the
actual form of  (\ref{psi.eq}),  and that in a generic  four-dimensional
spherically symmetric spacetime there may only be four independent equations
derived from the Einstein's equation (\ref{einst}).  In particular, equations
(\ref{Omega_tautau.0.eq}) and (\ref{r_tautau.0.eq}) can be seen to be
equivalent to the combinations
$E^\vartheta_{\ \vartheta}+E^\varphi_{\ \varphi}-E^\tau_{\ \tau}-E^\rho_{\ \rho}=0$
and $E^\tau_{\ \tau}+E^\rho_{\ \rho}=0$, whereas the
constraints are $E^\tau_{\ \rho}=0$ and $E^\tau_{\ \tau}=0$.

Notice that we have no evolution equation for $\beta$. According to the
terminology introduced by Friedrich in \cite{Friedrich}, this function can be
seen to play a role analogous to the ``gauge source function''. It is freely
specifiable, provided that it is sufficiently regular on the domain of time
evolution.  By making use of this freedom, we were able to slow down evolution
in our numerical simulations at the parts of the time level surfaces whenever
they have got close to the singularity. 

Defining the coordinate speed of light, $c_{[\tau,\rho]}=\drho/\dtau$ along
the radial null geodesics in the $\tau-\rho$ part---requiring the vanishing of
the pertinent part of the line element---, it is straightforward to see that
$c_{[\tau,\rho]}=\beta$. Thereby, as far as $\beta$ is guaranteed to be
bounded so is $c_{[\tau,\rho]}$. In our simulations, $\beta$ was guaranteed to
be less than or equal to 1.2 everywhere. It is also remarkable
that the system of evolution equations may
be viewed as a system of coupled non-linear wave equations for the
variables $\gotR$, $r$ and $\psi$. These type of equations are known to
possess a well-posed  initial value problem (see, e.g., \cite{ch}). In
addition, it can be checked by a direct calculation that whenever the
evolution equations are satisfied---and, in turn, their derivatives also
vanish---the following set of first order strongly hyperbolic evolution
equations can be derived
\begin{eqnarray}
&&\hskip-.6cm\partial_\tau E^\tau_{\ \rho}-\partial_\rho
  E^\tau_{\ \tau}+\frac2r\left[r_\tau E^\tau_{\ \rho}-r_\rho 
    E^\tau_{\ \tau}\right]+\frac{(\gotR_\tau\beta+\gotR\beta_\tau)
    E^\tau_{\ \rho}-(2\gotR_\rho\beta+\gotR\beta_\rho) 
    E^\tau_{\ \tau}}{\gotR\beta} \ =\ 0,\label{Cev1}\\ 
&&\hskip-.6cm\partial_\tau E^\tau_{\ \tau}-\beta^2\partial_\rho
  E^\tau_{\ \rho}+\frac2r\left[r_\tau E^\tau_{\ \tau}-\beta^2r_\rho 
    E^\tau_{\ \rho}\right]+\frac{\gotR_\tau E^\tau_{\ \tau}-
    \beta(\gotR_\rho\beta+3\gotR\beta_\rho) 
    E^\tau_{\ \rho}}{\gotR}\ =\ 0\,. \label{Cev2}
\end{eqnarray}
Since these equations are linear and homogeneous in the variables
$E^\tau_{\ \tau}$ and $E^\tau_{\ \rho}$, they ensure that the
constraints propagate with the 
evolution, therefore it is enough to impose them only on the initial data
hypersurface.


\medskip

In spite of the attractive features of the above outlined evolutionary
system---see equations (\ref{Omega_tautau.0.eq}-\ref{psi_tautau.eq}) and
(\ref{constraints.eq}) along with (\ref{Cev1})--(\ref{Cev2})---, it has been
found to be unstable in numerical simulation because truncation errors grow
rapidly near the origin(s), i.e., where $r$ tends to zero.  The instability is
caused by terms of the form
\begin{equation}
\frac{\gotR\beta^2+r_\tau^2 - \beta^2\,r_\rho^2}{r^n},\label{crashing_term.eq}
\end{equation}
appearing in (\ref{r_tautau.0.eq}) with $n=1$ and in (\ref{Omega_tautau.0.eq})
with $n=2$. The nominator of (\ref{crashing_term.eq}) consists of three
independent variables, each of which  contributes to its numerical error.
Although it should vanish rapidly while approaching the origin, it does not
exactly do that in practice and, in turn, its slightest error is  amplified
enormously via the division by $r^n$.

Increasing the resolution does not help in such a circumstance, the
simulations crash even earlier.  We tried to cure this instability in various
ways such as extrapolation from outer points, decreasing the order of finite
differences near the origin, increasing the numerical dissipation term, and
also by using different parametrisations of the metric which were supposed to
behave better close to the critical points.  Neither of these attempts turned
out to be successful. 

\section{Stabilising with the Misner-Sharp mass}\label{intr2}

By the inspection of the term (\ref{crashing_term.eq}) responsible for the
above mentioned instabilities and of the Misner-Sharp mass $m$
\cite{MisnerSharp1} defined by the relation 
\begin{equation}
m\ =\ \frac{r}{2}\left(1+g^{ab}\partial_a r\,\partial_b r\right)=
\frac{r}{2}\left(\frac{\gotR\beta^2+{r_\tau^2}-\beta^2\,r_\rho^2}
     {\gotR\beta^2}\right),
\label{m.def.eq}
\end{equation}
it seems to be rewarding to introduce $m$ as an auxiliary variable.  By
deriving it with respect to $\tau$ and $\rho$ and by using the pertinent form
of Einstein's equations (\ref{einst}) to eliminate the second derivatives of
$r$, an evolution and a constraint equation can be deduced:
\begin{align}
\partial_\tau m\ &=\,
\frac{r^2}{2}\left\{r_\tau\,E^\rho_{\ \rho}-r_\rho\,E^\rho_{\ \tau} 
+8\pi\,(r_\tau\,T^\rho_{\ \rho}-r_\rho\,T^\rho_{\ \tau})\right\}\,,
\label{m.prop1.eq}\\
\partial_\rho m\ &=\,
\frac{r^2}{2}\left\{r_\rho\,E^\tau_{\ \tau}-r_\tau\,E^\tau_{\ \rho} 
+8\pi\,(r_\rho\,T^\tau_{\ \tau}-r_\tau\,T^\tau_{\ \rho})\right\}\,.
\label{m.prop2.eq}
\end{align}
What makes the introduction of $m$ even more preferable is related to the
following observations. First of all, Einstein's equations $E_{ab}=0$
can be seen to hold whenever
\begin{align}
\partial_\tau\gotR_\tau\, &=\ \beta^2\partial_\rho\gotR_\rho 	+
\frac{4m\beta^2\gotR^2}{r^3} 
	+ 8\pi \gotR^2\beta^2\left(T^\vartheta_{\ \vartheta} 
+T^\varphi_{\ \varphi}-T^\tau_{\ \tau}-T^\rho_{\ \rho}\right) 
        \nonumber\\ &\hskip3.7cm+ \frac{\gotR_\tau^2 -
          \beta^2\gotR_\rho^2}{\gotR} + \frac{\beta_\tau 
          \gotR_\tau + \beta^2\beta_\rho \gotR_\rho}{\beta} +2\gotR\beta
        \partial_\rho\beta_\rho 
	\,,\label{Omega_tautau.2.eq}\\
\partial_\tau r_\tau\, &=\ \beta^2\partial_\rho r_\rho + 4\pi r\gotR
  \beta^2\,\left(T^\tau_{\ \tau} +  
  T^\rho_{\ \rho}\right) 
	- {2m\beta^2\gotR\over r^2}
	+ \frac{\beta_\tau r_\tau + \beta^2\beta_\rho r_\rho}{\beta},
	\label{r_tautau.2.eq}\\
\partial_\tau m\, &=\ 4\pi r^2 \left(r_\tau\,T^\rho_{\ \rho}
	- r_\rho\,T^\rho_{\ \tau}\right),\label{m_tau.2.eq}\\
\partial_\rho m\, &=\ 4\pi r^2 \left(r_\rho\,T^\tau_{\ \tau}-r_\tau\,T^\tau_{\ \rho}
\right).
\label{m_rho.2.eq}
\end{align}
It is important to emphasise that instead of solving the constraint equations
$E^\tau_{\ \tau}=0$ and $E^\tau_{\ \rho}=0$ on the initial data surfaces, it
suffices to solve (\ref{m_rho.2.eq}), provided that the evolution equations
(\ref{Omega_tautau.2.eq})-(\ref{m_tau.2.eq}) hold.  To see this, recall first
that (\ref{r_tautau.2.eq}) is equivalent to
\begin{equation}
E^\tau_{\ \tau}+E^\rho_{\ \rho}\ =\ 0.\label{seg1.eq}
\end{equation}
In virtue of (\ref{m.prop1.eq}), (\ref{m.prop2.eq}), (\ref{m_tau.2.eq}) and
(\ref{m_rho.2.eq}), we have that
\begin{equation}
r_\tau E^\rho_{\ \rho}-r_\rho E^\rho_{\ \tau}\ =\ 0,
\qquad r_\rho E^\tau_{\ \tau}-r_\tau E^\tau_{\ \rho}\ =\ 0.\label{seg2.eq}
\end{equation}
Equations (\ref{seg1.eq})-(\ref{seg2.eq}) along with the relation $E^\rho_{\
\tau} = -\beta^2 E^\tau_{\ \rho}$, imply then that for the variables
$E^\tau_{\ \tau}$, $E^\tau_{\ \rho}$, the homogeneous linear
equations
\begin{equation}
r_\tau E^\tau_{\ \tau} - \beta^2 r_\rho E^\tau_{\ \rho}\ =\ 0,
\qquad r_\rho E^\tau_{\ \tau} - r_\tau E^\tau_{\ \rho}\ =\ 0\label{seg3.eq}
\end{equation}
hold. The algebraic sub-determinant of this system is $r_\tau^2-\beta^2
r_\rho^2$, which will be shown in Section \ref{trap} to vanish only at
isolated marginally trapped surfaces. Therefore the system (\ref{seg3.eq})
possesses only the trivial solution which justifies our claim.

\medskip
Restricting our considerations to the case of a self-interacting scalar
field again, the components of the energy-momentum tensor can be
given as  
\begin{eqnarray}
T^\tau_{\ \tau}
&=& \frac{1}{2}\frac{\psi_\tau^2+\beta^2\psi_\rho^2}{\gotR\beta^2} + V,\\  
T^\tau_{\ \rho}\,\ =\,\ -\frac{1}{\beta^2}T^\rho_{\ \tau} &=&
\frac{\psi_\rho\psi_\tau}{\gotR\beta^2},\\ 
T^\rho_{\ \rho}
&=& -\frac{1}{2}\frac{\psi_\tau^2+\beta^2\psi_\rho^2}{\gotR\beta^2} + V,\\
T^\theta_{\ \theta}\,\ =\,\ T^\varphi_{\ \varphi} &=&
-\frac{1}{2}\frac{\psi_\tau^2-\beta^2\psi_\rho^2} 
	{\gotR\beta^2} + V.
\end{eqnarray}
In this case, the complete set of evolution equations consists of
(\ref{Omega_tautau.2.eq})-(\ref{m_tau.2.eq}) and (\ref{psi_tautau.eq}). 

\smallskip

Note that the introduction of $m$ makes the use of (\ref{Omega_tautau.2.eq})
superfluous since $\gotR$ can be determined from (\ref{m.def.eq}) once the
fields $m$ and $r$ are known, However, this way we get a 0/0 type formula for
$\alpha$ at `marginally trapped surfaces' (for their precise definition
see Section\,\ref{trap}) and an instability in the numerical simulation,
thereby making it impossible to study the evolution inside the black hole
region.  In order to overcome these difficulties, we kept
(\ref{Omega_tautau.2.eq}) as an evolution equation for $\gotR$. Then, as it
will be demonstrated in Section\,\ref{dyn} (see Fig.\,\ref{coll1}) we could
guarantee that our numerical framework does possess the desired capabilities
without the loss of numerical accuracy anywhere in the domain of dependence.

\medskip

By making use of standard procedures (described in details, e.g., in
\cite{ch}),  a first order evolutionary system for our spherically symmetric
dynamical system can be deduced as follows. In addition to $m$, $\gotR$, $r$,
and $\psi$, the first derivatives $\gotR_\tau$, $\gotR_\rho$, $r_\tau$,
$r_\rho$, $\psi_\tau$ and $\psi_\rho$ are also considered as being independent
variables, and the evolution equations are complemented by the relations
\begin{eqnarray}
\partial_\tau \gotR &=& \gotR_\tau,\\ \partial_\tau r&=& r_\tau,\\
\partial_\tau \psi&=& \psi_\tau,\\
\partial_\tau \gotR_\rho&=& \partial_\rho \gotR_\tau,\\
\partial_\tau r_\rho&=&\partial_\rho r_\tau,\\
\partial_\tau \psi_\rho&=&\partial_\rho \psi_\tau\,.
\end{eqnarray}
Note that the last three
relations are, in fact, the integrability conditions for $\gotR$, $r$, and
$\psi$. As a byproduct of this reduction process, in addition to the true
dynamical constraint (\ref{m_rho.2.eq}),
we also have to take care of the trivial ones 
\begin{eqnarray}
\partial_\rho \gotR &=& \gotR_\rho,\label{cO}\\ \partial_\rho r&=&
r_\rho,\label{cr}\\ 
\partial_\rho \psi&=& \psi_\rho\,\label{cP}.
\end{eqnarray}  
The yielded first order system can then be put into the form 
\begin{equation}
\partial_\tau u\,\ =\,\ {A}\,\partial_\rho u + B,
\label{u_tau.0.eq}
\end{equation}
where the associated $10$-dimensional vector variable $u$, the $10\times10$
matrix $A$ and the $10$-dimensional source vector $B$ are given as 
\begin{eqnarray}
u\ =\ \left(\begin{matrix}
	m\cr \gotR\cr \gotR_\tau\cr \gotR_\rho\cr
	r\cr r_\tau\cr r_\rho\cr
	\psi\cr \psi_\tau\cr \psi_\rho \end{matrix}\right),
\qquad A \ =\ \left(\begin{matrix}
	0 & 0 & 0 & 0 & 0 & 0 & 0 & 0 & 0 & 0\cr
        0 & 0 & 0 & 0 & 0 & 0 & 0 & 0 & 0 & 0\cr
	0 & 0 & 0 & \beta^2 & 0 & 0 & 0 & 0 & 0 & 0\cr
	0 & 0 & 1 & 0 & 0 & 0 & 0 & 0 & 0 & 0\cr
	0 & 0 & 0 & 0 & 0 & 0 & 0 & 0 & 0 & 0\cr
	0 & 0 & 0 & 0 & 0 & 0 & \beta^2 & 0 & 0 & 0\cr
	0 & 0 & 0 & 0 & 0 & 1 & 0 & 0 & 0 & 0\cr
	0 & 0 & 0 & 0 & 0 & 0 & 0 & 0 & 0 & 0\cr
	0 & 0 & 0 & 0 & 0 & 0 & 0 & 0 & 0 & \beta^2\cr
	0 & 0 & 0 & 0 & 0 & 0 & 0 & 0 & 1 & 0\cr
	\end{matrix}\right),
\end{eqnarray}
\begin{eqnarray}
B \ =\ \left(\begin{matrix}
        -\frac{2\pi r^2}{\gotR\beta^2}
		\left(r_\tau (\psi_\tau^2+\beta^2\psi_\rho^2)
		- 2r_\rho\beta^2\psi_\tau\psi_\rho\right)
                + 4\pi r^2 r_\tau V\cr
	\gotR_\tau\cr
	\frac{\gotR_\tau^2 - \beta^2\gotR_\rho^2}{\gotR}
	+ \frac{\alpha_\tau\beta_\tau+\beta^2\alpha_\rho\beta_\rho}{\beta}
	+ 2\gotR\beta\,\partial_\rho\beta_\rho
        - 8\pi \gotR\beta^2(\psi_\tau^2 - \beta^2\psi_\rho^2)
        + \frac{4m\gotR^2\beta^2}{r^3}\cr
	0\cr
	r_\tau\cr
	- \frac{2m\gotR\beta^2}{r^2}+ \frac{\beta_\tau r_\tau + \beta^2\beta_\rho
          r_\rho}{\beta} +8\pi r\gotR \beta^2\,V \cr
	0\cr
	\psi_\tau\cr
- 2\frac{\psi_\tau r_\tau - \beta^2\psi_\rho r_\rho}{r}
	+ \frac{\beta_\tau\psi_\tau + \beta^2\beta_\rho\psi_\rho}{\beta}
	- \gotR\beta^2\frac{\partial V}{\partial\psi}\cr
	0\cr
	\end{matrix}\right)\,\label{B}.
\end{eqnarray}
Since the eigenvectors of the matrix $A$ comprise a complete system and its
eigenvalues are all real, this first order system is strongly
hyperbolic \cite{gko}.\footnote{%
  Note that defining $\hat\gotR_\rho=\beta\gotR_\rho$,
  $\hat r_\rho=\beta r_\rho$ and $\hat\psi_\rho=\beta\psi_\rho$, the above set
  of field equations could be put into the form of a first order symmetric
  hyperbolic system for the vector valued variable
  $(m,\gotR,\gotR_\tau,\hat\gotR_\rho,r,r_\tau,\hat r_\rho,
  \psi,\psi_\tau,\hat \psi_\rho)^T$.}
It is known that both the analytic and numerical well-posedness of the initial
value problem is guaranteed for this type of evolution equations
\cite{ReulaHypMethodsForEeqs,gko}.  An additional preferable property of the
investigated dynamical system is that for the coordinate speed of light,
$c_{[\tau,\rho]}\leq \beta_\mathrm{max}$ holds, where $\beta_\mathrm{max}$
denotes the maximum value of $\beta$.

\section{Stability in the origin(s)}\label{stabor} 

The behaviour of the basic field variables $m,\gotR, \beta, r, \psi$ close to an origin located, say, at
$\rho=\rho_0$ can be explored by substituting the form of Taylor expansion
\begin{equation}
f(\tau,\rho) = \sum_{k=0}^\ell \frac{1}{k!}
f_k (\tau)\, (\rho-\rho_0)^k
+\mathcal{O}_f((\rho-\rho_0)^{\ell+1}),\label{taylor}
\end{equation}
into the field equations for the functions $m,\gotR, \beta, r, \psi$, and also
by
requiring the solutions to be at least of class $C^\ell$ for some
$\ell\in\mathbb{N}$.  Assuming that the solutions are at least of class $C^4$
in a neighbourhood of $\rho_0$, the resulting relations are
\begin{eqnarray}
&&m=
{\displaystyle \frac {1}{6}} \,{\mathit{m}_{3}}\,(\rho-\rho_0) ^{3} + 
\mathcal{O}_m((\rho-\rho_0) ^{5})\label{reg2_m}\\
&&\gotR ={\gotR _{0}} + {\displaystyle \frac {1}{2}} \,{
\gotR _{2}}\,(\rho-\rho_0) ^{2} + {\displaystyle \frac {1}{24}} \,{
\gotR _{4}}\,(\rho-\rho_0) ^{4} + \mathcal{O}_\gotR((\rho-\rho_0)
^{6}) \label{reg2_Omega}\\ 
&&\beta ={\beta_{0}} + {\displaystyle \frac {1}{2}} \,{
\beta_{2}}\,(\rho-\rho_0) ^{2} + {\displaystyle \frac {1}{24}} \,{
\beta_{4}}\,(\rho-\rho_0) ^{4} + \mathcal{O}_\beta((\rho-\rho_0)
^{6}) \label{reg2_beta}\\
&&r=\sgn(\rho-\rho_0)\sqrt{{\gotR _{0}}}\,(\rho-\rho_0)  + 
{\displaystyle \frac {1}{6}} \,{\mathit{r}_{3}}\,(\rho-\rho_0) ^{3} +
\mathcal{O}_r((\rho-\rho_0) ^{5})\label{reg2_r} \\
&&\psi ={\psi _{0}} + {\displaystyle \frac {1}{2}} \,{\psi _{
2}}\,(\rho-\rho_0) ^{2} + {\displaystyle \frac {1}{24}} \,{\psi _{4}}\,
(\rho-\rho_0) ^{4} + \mathcal{O}_\psi((\rho-\rho_0) ^{6})\,.\label{reg2_psi}
\end{eqnarray}

By making use of the indicated parity properties of the functions
$\gotR,\beta,r$ and $\psi$, the grid boundary at the origin can be treated
exactly the same way as it was done in \cite{FR,FR2}. In particular, the
method described in details in Section D of \cite{FR2} makes it possible to
determine all the spatial derivatives in a neighbourhood of an origin, by
making use of a symmetric stencil.

\smallskip

Note that the factor $\sgn(\rho-\rho_0)\sqrt{\gotR_0}$ in the first order term
in (\ref{reg2_r}) has a simple geometrical explanation. In order to have a
regular origin, i.e., to avoid the appearance of  conical singularities, we
have to guarantee that while approaching to the origin, $\rho\rightarrow
\rho_0$, the ratio of the circumference of an infinitesimal, origin-centred
circle and of the corresponding proper radius tends to $2\pi$, i.e.,
\begin{equation}
\frac{2\pi r}{|\int_{\rho_0}^\rho\hskip-.1cm\sqrt{\gotR}\,\drho|}
\ \rightarrow\ \frac{2\pi\partial_\rho r}{\sgn(\rho-\rho_0)\sqrt{\alpha}}
 \ \rightarrow\ 2\pi,
\end{equation}
where l'Hospital's rule is also applied.

\medskip

Having the relations (\ref{reg2_m}) and (\ref{reg2_r}), it is also
straightforward to see why the system introduced in Section\,\ref{intr2} has
been found stable in our numerical simulations. One of the reasons
for this unexpected  stability is related to the fact that the term
(\ref{crashing_term.eq})---which was responsible for the numerical
instabilities in case of the simple-minded system---reads as  
\begin{equation}
\frac{\gotR\beta^2+r_\tau^2-\beta^2r_\rho^2}{r^n}\ =\ \frac{2\,m\,\gotR\beta^2}
     {r^{n+1}}\ =\ \frac{m_3\,\gotR\beta^2}{3\alpha^{(n+1)/2}}
  |\rho-\rho_0|^{2-n}\,. 
\label{crashing_term2.eq} 
\end{equation}
Since $n$ takes the values $1$ and $2$, the term (\ref{crashing_term.eq}) is
replaced by a completely regular one in our new setup. It is also important to
note that the Misner-Sharp mass $m$
is subject to a first order PDE (\ref{m_tau.2.eq}) which does not involve a
singular term at all. 

\medskip 

If there is an origin, it is useful to have the limiting form of the source
vector $B$ appearing in (\ref{u_tau.0.eq}). By making use of
(\ref{reg2_m})-(\ref{reg2_psi}), it is straightforward to verify that the
non-trivial  components $B_{\gotR_\tau}$, $B_{r_\tau}$ and $B_{\psi_\tau}$ of
$B$, as given in (\ref{B}), tend to the following regular limits 
\begin{align}
\lim_{\rho\rightarrow\rho_0} B_{\gotR_\tau}
	&=\,\frac32\frac{(\partial_\tau\gotR_0)^2}{\gotR_0}
        +\frac{(\partial_\tau\gotR_0)(\partial_\tau\beta_0)}{\beta_0} 
	- 2 \sgn(\rho-\rho_0)\sqrt{\gotR_0}\,\beta_0^2\,r_3+\gotR_2\,\beta_0^2
        \nonumber\\ & \phantom{\,\frac32\frac{(\partial_\tau\gotR_0)^2}{\gotR_0}
        +\frac{(\partial_\tau\gotR_0)(\partial_\tau\beta_0)}{\beta_0} 
	\sgn(\rho-\rho_0)\sqrt{\gotR_0}}+2\gotR_0\beta_0\beta_2 
        -8\pi\gotR_0 (\partial_\tau\psi_0)^2\,,\\ 
\lim_{\rho\rightarrow\rho_0}B_{r_\tau}
	&=\, 0\,,\\
\lim_{\rho\rightarrow\rho_0}B_{\psi_\tau}
	&=\,
        -\frac{(\partial_\tau\psi_0)(\partial_\tau\gotR_0)}{\gotR_0}
        +2\beta_0^2\,\psi_2 - \gotR_0\beta_0^2\,\frac{\partial
          V}{\partial\psi}\,.
\end{align}

\section{Energy balance}\label{energybal}

As it is well-known, energy transfer processes cannot be investigated in
general relativity, since appropriate quasi-local quantities have not been
found as yet. However, this can be done in spherically symmetric dynamical
spacetimes, using the Kodama vector field
\cite{kodama}
\begin{equation}
K^a\ =\ \epsilon^{ea}{\partial_e}r, \label{kv0} 
\end{equation}
where $\epsilon^{ab}=q^{ae}q^{bf}\epsilon_{ef}$ and $\epsilon_{ab}$ denotes
the volume form associated with the metric $q_{ab}$ induced on the
two-dimensional surface transverse to the $SO(3)$ group orbits. It was shown
in \cite{kodama} that $K^a$ is divergence free, i.e., $\nabla^e K_e=0$, and
also that $G^{ef}\nabla_e K_f$ vanishes, where $G_{ab}$ denotes the Einstein
tensor $R_{ab}-\frac12 Rg_{ab}$. It follows then that the vector field 
\begin{equation}\label{kv00}
J^a\ =\ {T^{a}}_{b}K^b
\end{equation}
is also divergence free. Accordingly, even though our spherically symmetric
spacetime is fully dynamical, $J^a$ behaves as a locally conserved energy flux
type vector field. Then, by making use of Stokes' theorem and by choosing $Q$
to be a four dimensional spacetime region with boundary $\partial Q$, we have
that 
\begin{equation}
\int_Q\nabla_aJ^a\epsilon\ =\ \int_{\partial Q}n_aJ^a\tilde
\epsilon\ =\ 0\,, \label{stokes}
\end{equation}
where $\epsilon$ denotes the 4-volume element while $\tilde \epsilon$ is the
3-volume element induced on the boundary $\partial Q$ of $Q$, which can be
given as  $\tilde \epsilon_{abc}={\epsilon}_{eabc}n^e$, where $n_a$ is the
(outward pointing) unit normal 1-form field on $\partial Q$. 

With our particular choice of coordinates, the components of the Kodama vector
field can be given as 
\begin{equation}
K^\alpha\ \,=\ \, \left(\frac{r_\rho}{\gotR\beta},
\ -\frac{r_\tau}{\gotR\beta},\ 0,\ 0\right)\,,
\end{equation}
which, in virtue of (\ref{kv00}), implies that the non-zero components of the
divergence free vector field $J^a$ read as 
\begin{equation}\label{Jt}
J^\tau=\frac{1}{\gotR\beta}\left[T^\tau_{\ \tau}r_\rho -
  T^\tau_{\ \rho}r_\tau\right]\ \ \ {\rm and }\ \ \
J^\rho=\frac{1}{\gotR\beta}\left[T^\rho_{\ \tau}r_\rho -
  T^\rho_{\ \rho}r_\tau\right]\,.
\end{equation}

\begin{figure}[ht]
\unitlength1cm
 \centerline{
  \epsfysize=6.5cm 
  \epsfbox{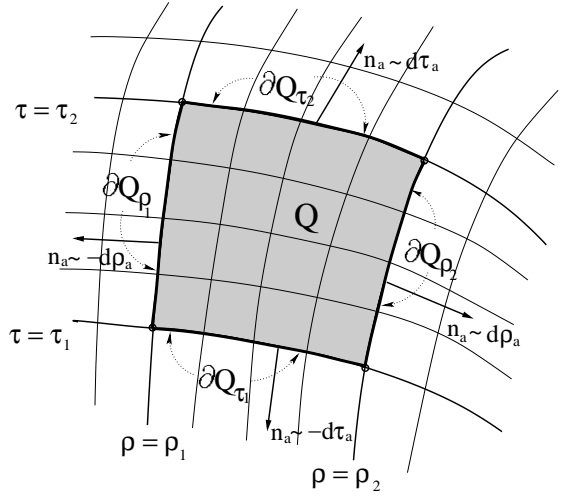}
 }
\caption{\footnotesize \label{enbal}   {The boundary $\partial Q=\partial
    Q_{\tau_2}\cup\partial Q_{\rho_2}\cup\partial Q_{\tau_1}\cup\partial
    Q_{\rho_1}$  of the shaded spacetime domain, enclosed by the
    $\tau=\tau_1$, $\tau=\tau_2$, $\rho=\rho_1$ and $\rho=\rho_2$
    hypersurfaces. }}  
\end{figure}

To characterise the energy transfer processes in a spacetime domain $Q$
bounded by sections of  $\tau=const$ and $\rho=const$ hypersurfaces, as it is
depicted on Fig.\,\ref{enbal}, we introduce the following auxiliary notations.

Denote the origin-centred ball of radius $\bar\rho$ on the $\tau=\bar \tau$
hypersurface by $\mathcal{B}(\bar \tau,\bar \rho)$, and the part of the
cylindrical hypersurface $\rho=\bar \rho$ between $\tau=\tau_1$ and $\tau_2$
by $\mathcal{C}(\tau_2,\tau_1,\bar \rho)$.  Then the boundaries $\partial
Q_{\tau_i}$ ($i=1,2$), as shown on Fig.\,\ref{enbal}, are represented by the
differences
$\Delta\mathcal{B}(\tau_i,\rho_2,\rho_1)=\mathcal{B}(\tau_i,\rho_2) \setminus
\mathcal{B}(\tau_i,\rho_1)$ of the balls $\mathcal{B}(\tau_i,\rho_2)$ and
$\mathcal{B}(\tau_i,\rho_1)$, while $\partial Q_{\rho_i}$ are the cylinders
$\mathcal{C}(\tau_2,\tau_1,\rho_i)$ connecting the boundaries of the balls
$\mathcal{B}(\tau_2,\rho_i)$ and $\mathcal{B}(\tau_1,\rho_i)$. 

Then the energy contained in $\Delta\mathcal{B}(\bar \tau,\rho_2,\rho_1)$ 
can be given as 
\begin{equation}\label{sint}
E(\bar \tau;\rho_2,\rho_1)\ =\ \int_{\Delta\mathcal{B}(\bar \tau,\rho_2,\rho_1)}
n_a^{(\bar \tau)}J^a\tilde\epsilon^{\,(\bar \tau)} \,, 
\end{equation}
where $\tilde{\epsilon}^{\,(\bar \tau)}$ is the volume element on the $\tau=\bar
\tau$ hypersurface and $n_a^{(\bar \tau)}$ is its unit normal $1$-form
field, 
i.e.{,}
\begin{equation} 
n_a^{(\bar\tau)}\ =\ \frac{\nabla_a \tau}{\sqrt{|(\nabla^e \tau)
    (\nabla_e \tau)}|}\Big\vert_{\tau=\bar \tau} \,.
\end{equation}
Similarly, the energy transported through the cylindrical
hypersurface $\mathcal{C}(\tau_2,\tau_1,\bar \rho)$ is
\begin{equation}\label{tint}
S(\tau_2,\tau_1;\bar \rho)\ =\ \int_{\mathcal{C}(\tau_2,\tau_1,\bar \rho)}
n_a^{(\bar\rho)}J^a \tilde\epsilon^{\,(\bar\rho)} \,, 
\end{equation}
where $\tilde\epsilon^{\,(\bar\rho)}$ is the volume element on the
$\rho=\bar\rho$ hypersurface and $n_a^{(\bar\rho)}$ is its unit normal
$1$-form field, i.e.{,}
\begin{equation} 
n_a^{(\bar\rho)}\ =\ \frac{\nabla_a \rho}{\sqrt{|(\nabla^e \rho)
    (\nabla_e \rho)}|}\Big\vert_{\rho=\bar \rho} \,.
\end{equation} 
Using these notations, the energy balance equation (\ref{stokes})
takes the form
\begin{equation}\label{stokes2}
E(\tau_2;\rho_2,\rho_1)-E(\tau_1;\rho_2,\rho_1)
+S(\tau_2,\tau_1;\rho_2)-S(\tau_2,\tau_1;\rho_1)\ =\ 0\,. 
\end{equation}
The following relations can be verified by straightforward calculation:
\begin{align} 
n_a^{(\tau)}\ &=\ (\sqrt{\gotR}\,\beta,0,0,0),\qquad
n_a^{(\rho)}\ =\ (0,\sqrt{\gotR},0,0)\,, \\
\tilde \epsilon_{\alpha\beta\gamma}^{\,(\tau)}\ &=\ \sqrt{\gotR}\,
r^2\sin\theta\cdot(\drho)_{\alpha}\wedge(\dtheta)_\beta\wedge(\dvarphi)_\gamma\,,
\label{vol_tau.eq}\\
\tilde \epsilon_{\alpha\beta\gamma}^{\,(\rho)}\ &=\ \sqrt{\gotR}\,\beta\,
r^2\sin\theta\cdot(\dtau)_{\alpha}\wedge(\dtheta)_\beta\wedge(\dvarphi)_\gamma\,.
\label{vol_rho.eq}
\end{align} 
Thus $E(\bar\tau;\rho_2,\rho_1)$ and $S(\tau_2,\tau_1;\bar\rho)$ can be given
as
\begin{align}
E(\bar\tau;\rho_2,\rho_1)\ &=\ 4\pi\int_{\rho_1}^{\rho_2}
(\gotR\,\beta\,r^2\,J^\tau)|_{\tau=\bar\tau}\, \drho ,\\
S(\tau_2,\tau_1;\bar\rho)\ &=\ 4\pi\int_{\tau_1}^{\tau_2}
(\gotR\,\beta\,r^2\,J^\rho)|_{\rho=\bar\rho}\, \dtau\,. 
\end{align}

It is important to mention here that, in virtue of the relations
(\ref{m_rho.2.eq}), (\ref{Jt}) and (\ref{vol_tau.eq})-(\ref{vol_rho.eq}), the
Misner-Sharp mass function $m$ can be given on a $\tau=\bar \tau$ time level
surface as
\begin{equation}\label{mtau}
m(\bar \tau,\rho)\ =\ E(\bar\tau;\rho,\rho_1=0)
\ =\ m(\bar \tau,0)+4\pi\int_{0}^{\rho}
(\gotR\,\beta\,r^2\,J^\tau)|_{\tau=\bar\tau}\, \drho \,.
\end{equation}
Accordingly, in virtue of (\ref{vol_tau.eq})-(\ref{vol_rho.eq}) and
(\ref{mtau}), the expression $\varepsilon_{_{GM}}=
\sqrt{\gotR}\,\beta\,J^\tau$ could also be interpreted as the combined energy
density of our composed gravity-matter system.

\section{Trapped surfaces}\label{trap}

To define trapped surfaces, start with a
smooth orientable $2$-dimensional compact manifold $\mathscr{S}$ with no
boundary in a $4$-dimensional spacetime $(M,g_{ab})$. Let $n^a$ be a smooth
future directed non-vanishing null vector field on $\mathscr{S}$ which is
normal to $\mathscr{S}$, i.e., $g_{ab}n^aX^b|_{\mathscr{S}}=0$ for any vector
field  $X^a$ tangent to $\mathscr{S}$.  Consider then the null hypersurface
$\mathcal{N}$ generated by geodesics starting on $\mathscr{S}$ with  tangent
  $n^a$. By parallel propagating $n^a$ along these generators, it can be
extended onto $\mathcal{N}$. Denote by $u$ the synchronised affine
parameterisations of these null 
  geodesics. The
  hypersurface $\mathcal{N}$ is  smooth in a  neighbourhood of $\mathscr{S}$,
  and it is smoothly foliated by the $u$-level surfaces.
  Denote by ${\eeepsilon}_q$ the volume
  element associated to the metric $q_{ab}$ induced on these
  $2$-dimensional surfaces. Then the null expansion $\theta^{(n)}$ with
  respect to  $n^a$ is defined as
\begin{equation}\label{exp}
\pounds_n\,{\eeepsilon}_q= \theta^{(n)}\,{\eeepsilon}_q\,,   
\end{equation}  
where $\pounds_n$ denotes the Lie derivative with respect to the null vector
field $n^a$.  

It is straightforward to see that the sign of $\theta^{(n)}$ remains intact
under a positive rescaling of  $n^a$, i.e., whenever $n^a$ is replaced by
$n'^a=f\,n^a$, where $f$ is a sufficiently smooth positive function on
$\mathscr{S}$. Therefore, as far as the sign of $\theta^{(n)}$ is concerned,
$n^a$ and  $n'^a$ may be considered to be equivalent.  It is well-known that
there always exist two equivalence classes of future directed  non-vanishing
null vector fields on $\mathscr{S}$ which are normal to $\mathscr{S}$.  Let
$n^a_+$ and $n^a_-$ be two such vector fields and denote the corresponding
null expansions by $\theta_{+}$ and $\theta_{-}$.\footnote{
  Apart from the case of asymptotically flat spacetimes, the
  '$+$' and '$-$' signs have nothing to do with outward and inward pointing
  directions. Although these directions cannot be defined for generic
  spacetimes, it is worthwhile to mention that an adequate quasi-local notion of
  outward and inward direction can be introduced---without referring to global
  properties of the underlying spacetimes---provided that attention is
  restricted to untrapped or marginally trapped surfaces (see, e.g.
  \cite{Racz2}).}  According to the original definition of Penrose \cite{P}, a
$2$-dimensional surface $\mathscr{S}$ in a $4$-dimensional spacetime is {\it
  future trapped} or {\it untrapped} if both of the future directed null
geodesic congruences orthogonal to $\mathscr{S}$ are converging, or one of
them is diverging while the other is converging at $\mathscr{S}$, i.e.,
$\theta_{\pm}\leq 0$ or $\theta_{+}$ and $\theta_{-}$ have opposite signs. If
one of the expansions vanishes identically while the other is everywhere
non-positive, then $\mathscr{S}$ is called future {\it marginally trapped}
surface.  {\it Past trapped} and past {\it marginally trapped} surfaces can be
defined analogously by reversing the time orientation applied tacitly in the above
definitions. 

Choose now $\mathscr{S}$ to be a $2$-sphere of radius $r$, invariant
under the $SO(3)$ symmetry of our spherically symmetric spacetime. In virtue
of the $\rho-\tau$ part of the metric (\ref{the_metric.eq}), an obvious choice
for the non-vanishing future directed null vector fields $n^a_+$ and $n^a_-$
on $\mathscr{S}$ is
\begin{equation}\label{npnm}
n^a_\pm=\left(\frac{\partial}{\partial \tau}\right)^a \pm
\beta\left(\frac{\partial}{\partial \rho}\right)^a\,,
\end{equation}
while the $2$-volume element ${\eeepsilon}_q$ on the $2$-spheres
$\mathscr{S}_{\tau,\rho}$ foliating the $\mathcal{N}_\pm$ null hypersurfaces
is
\begin{equation}\label{eps}
{\mathbf{\eeepsilon}}_q=r^2\,\sin\vartheta\cdot{\bf e},
\end{equation}
where ${\bf e}$ denotes the Levi-Civita symbol. It is
straightforward to check that the pertinent null expansions are
\begin{equation}\label{thpm}
\theta_\pm=\frac2r\cdot\left(r_\tau\pm \beta r_\rho\right)\,.
\end{equation}
Equations (\ref{m.def.eq}) and (\ref{thpm}) imply that the following relation
holds on $\mathscr{S}$:
\begin{equation}\label{thpm2}
1-\frac{2m}{r}
\ =\ \frac{\beta^2r_\rho^2-r_\tau^2}{\gotR\beta^2}
\ =\ -\frac{r^2}{4\,\gotR\beta^2}\cdot\theta_+\,\theta_-\,.
\end{equation} 

Since all the functions $\gotR$, $\beta$ and $r^2$ are positive, apart from
origin(s) or the spacetime singularity, an $SO(3)$ invariant $2$-sphere
$\mathscr{S}$ of radius $r$ is untrapped or trapped (future or past) if and
only if $1-\frac{2m}{r}$ is positive or negative, respectively. 

We shall call a spacetime region $\mathscr{T}_+$ future trapped if each point
of it belongs to an $SO(3)$ invariant trapped 2-sphere. The past boundary,
$\partial^-\mathscr{T}_+ = \overline{\mathscr{T}}_+\cap J^-[\mathscr{T}_+]$,
of such a future trapped region will be referred as {\it future apparent
  horizon} $\mathscr{A}_+$.\footnote{ It is important to keep in mind that
  whenever non-invariant topological 2-spheres are involved, the extent of the
  trapped region might be larger than the one determined by $SO(3)$-invariant
  2-spheres \cite{Senovilla}.} The notion of past trapped region,
$\mathscr{T}_-$, and {\it past apparent horizon}, $\mathscr{A}_-$, may be
introduced analogously. 

\section{Dynamical investigations}\label{dyn}

The first order strongly hyperbolic system described in Section\,\ref{intr2}
and given in the form (\ref{u_tau.0.eq}) was solved numerically with
our finite difference code called GridRipper AMR \cite{GridRipper}
(see also \cite{csp,csp2}). The time integration process uses the
method of lines based on a fourth order Runge-Kutta scheme \cite{gko}---for a
detailed description of this method, for the case of fixed spatial
resolutions, see, e.g., \cite{FR2}---and as it is indicated by its name,
adaptive mesh refinement techniques are also incorporated.

\medskip

\subsection{Initial data}\label{inidata}

Before determining the time evolution of our system using equation
(\ref{u_tau.0.eq}), we have to specify suitable initial data on a spacelike
hypersurface $\Sigma_0$---which hereafter will always be chosen as the
$\tau=0$ hypersurface---so that all the  constraint equations are satisfied.
Recall that the components of the vector variable $u$ in
(\ref{u_tau.0.eq}) are
$m,\gotR,\gotR_\rho,\gotR_\tau,r,r_\rho,r_\tau,$$\psi,\psi_\rho$ and
$\psi_\tau$ and, as it follows from the discussion in Section\,\ref{intr2}, we
apparently have only the constraint
\begin{equation}\label{cm}
\partial_\rho m = \frac{2\pi r^2}{\gotR\beta^2}
\left[r_\rho\,(\psi_\tau^2+\beta^2\psi_\rho^2 + 2\gotR
  \beta^2V)-2\,r_\tau\,\psi_\tau\,\psi_\rho 
\right]
\end{equation}
beside the three trivial ones (\ref{cO})--(\ref{cP}). However, the price to be
paid for using the ``superfluous'' evolution equation
(\ref{Omega_tautau.2.eq}) is the appearance of an additional constraint
(\ref{m.def.eq}), which implies that only two of the four variables,
$\gotR,\gotR_\tau,r$ and $r_\tau,$ may be chosen freely.  To satisfy all
constraints on the initial hypersurface, we have applied either of the
following two selection procedures.

\begin{itemize}
\item[A.] Fixing of $r$ and $r_\tau$ from among the functions
$\gotR,\gotR_\tau,r$ and $r_\tau$ on $\Sigma_0$.
Start by specifying the sufficiently regular but otherwise arbitrary functions
$r,r_\tau,\psi$ and $\psi_\tau$, then choose $r_\rho$ and $\psi_\rho$ so as
to satisfy the trivial constraints (\ref{cr}) and (\ref{cP}). By
substituting the relation
\begin{equation}\label{Om}
\gotR=\left(1-\frac{2m}{r}\right)^{-1}\frac{\beta^2\,r_\rho^2-r_\tau^2}{\beta^2}
\end{equation}
into (\ref{cm}), the initial data for $m$ can be determined
numerically.  Then the initial $\gotR$ and $\gotR_\rho$ are determined by
(\ref{Om}) and the trivial constraint (\ref{cO}), respectively. Finally,
$\gotR_\tau$ is
the $\tau$-derivative of (\ref{Om}),
\begin{equation}
\gotR_\tau = 2\frac{(r\,m_\tau - m\,r_\tau)
	(\beta^2\,r_\rho^2 - r_\tau^2)
	+ r (r - 2m) \left(\beta^2\,r_\rho\cdot\partial_\rho r_\tau
			   - r_\tau\cdot\partial_\tau r_\tau
                           +r_\tau^2\,\beta_\tau/\beta\right)} 
	{\beta^2(r - 2m)^2}\,.\label{Omega_tau.initcond.eq}
\end{equation}

\item[B.] Fixing of $\gotR$, $r$, $\psi$ and $\psi_\tau$ on $\Sigma_0$.
Choose $\gotR_\rho,r_\rho$ and $\psi_\rho$ so as to satisfy the trivial
constraints (\ref{cO})-(\ref{cP}).  From (\ref{m.def.eq}), we get
\begin{equation}\label{rtau}
r_\tau=\pm\,\beta\,\sqrt{\,r_\rho^2-\gotR\left(1-\frac{2m}{r}\right)}\,,
\end{equation}
where the sign depends on the specific physical problem we intend to
investigate.  By substituting the right hand side of (\ref{rtau}) into
(\ref{cm}), the yielded equation can be solved (numerically) for $m$ on the
initial data surface $\Sigma_0$. Then the initial data for $r_\tau$ is
determined by the relation (\ref{rtau}). Finally, $\gotR_\tau$ can be
determined similarly as in case A.
\end{itemize}
Procedure A. is simpler but B. provides a convenient control of both the
temporal geometrical setup and the energy distribution of the matter field.
The latter is also more appropriate to investigations of expanding
cosmological models.  Other choices may also be possible, however, the use of
these two cases was found to be completely satisfactory. {It is worth to
  be emphasised that either of these initial data specifications
  is suitable to host all the possible Einstein-scalar field systems in the
  selected setup.} 

To summarise, we have free control only of the four dynamical variables
$r,r_\tau,\psi$ and $\psi_\tau$ or $\gotR,r,\psi$ and $\psi_\tau$ in the two
cases, respectively. In addition, the specification of $\beta$ is also
required.  Once these functions are specified, they determine all other
functions by either of the outlined procedures on $\Sigma_0$.  In our
numerical simulations, we always used $\beta=1$ on $\Sigma_0$.

\subsection{Dynamical lapse function}

As it was emphasised earlier, the lapse function $\beta$ is freely
specifiable. Also, as it was indicated by the simple example in the
introduction, the extent of the spacetime domain covered by the time level
surfaces is influenced significantly by the chosen lapse function.  To achieve
the largest possible domain, we use a dynamically determined $\beta$
satisfying the evolution equation
\begin{equation}
\beta_\tau=-p\left[ \frac{m_\tau\,r-3\,m\,r_\tau}{r^4}
  \right]\left(\frac{r_\tau}{r}\right)^3\beta \,,\label{beta_tau.eq}
\end{equation}
where $p$ is a positive real parameter, starting with the initial value
$\beta|_{\Sigma_0}=1$. The value of the parameter was chosen to be $p=10^{-4}$
in all of our simulations.

Equation (\ref{beta_tau.eq}) was motivated as follows.  To slow down the
evolution close to the developing singularities, we need a function which
tends to zero there rapidly enough. For this reason, first we applied
$\beta=\exp(-pm/r^3)$. Besides decaying fast enough, this function is also
regular at the origin since, in virtue of (\ref{reg2_m}) and (\ref{reg2_r}),
$m/r^3$ remains finite while $\rho\rightarrow\rho_0$. The $\tau$-derivative of
this $\beta$ is ``almost'' (\ref{beta_tau.eq}), but without the factor
$(r_\tau/r)^3$ on the right hand side. The inclusion of this factor was found
to be necessary in stabilising the time evolution close to the
singularity. {Its use is supported by the observation that in a
  neighbourhood of a  point where the function $r$ takes its minimum the
  factor $({r_\tau}/{r})^3$ gets to be sufficiently small providing thereby 
  some additional slowing down to the evolution of $\beta$ there.} 

\subsection{Gravitational collapse with spatial topology
  $\mathbb{R}^3=[0,\infty)\times\mathbb{S}^2$}\label{R3coll} 
 
Let us start by investigating the gravitational collapse of a massive real
scalar field in the ``conventional'' case, i.e., with the assumption that the
initial data is specified on a hypersurface $\Sigma_0$ possessing the
topology of $\mathbb{R}^3$. According to procedure A, we specify two metric
functions on $\Sigma_0$ as
\begin{equation}\label{ic1}
r(\rho)\ =\ \rho,\qquad r_\tau(\rho)\ =\ 0\,.
\end{equation}
The scalar field $\psi$ is chosen to be a smooth hunch with compact support
\begin{align}\label{ic2}
\psi(\rho)\ &=\ \begin{cases}
	c \exp\left(d + \frac{b^2 d}{(\rho-a)^2 - b^2}\right)
		& \mathrm{if}\ |\rho-a| < b,\cr
	0 & \mathrm{if}\ |\rho-a| \geq b,
	\end{cases}
\end{align}
with vanishing $\tau$-derivative on $\Sigma_0$ 
\begin{equation}\label{ic3}
\psi_\tau(\rho)\,=\,0.
\end{equation}
The self-interaction potential is assumed to possess the form
$V(\psi)=\frac{1}{2}\mu^2\psi^2$.
The parameter values are
\begin{equation}\label{ic4}
a\,=\,0,\quad b\,=\,70,\quad c\,=\,0.0795,\quad d\,=\,100,\quad  \mu\,=\,0.8\,.
\end{equation}
In the numerical simulation, AMR was used with $7,000$ spatial points on the
base grid and five refinement levels (refinement ratio 2).

\begin{figure}[h]
\unitlength1cm
 \centerline{
  \epsfysize=6.cm 
  \epsfbox{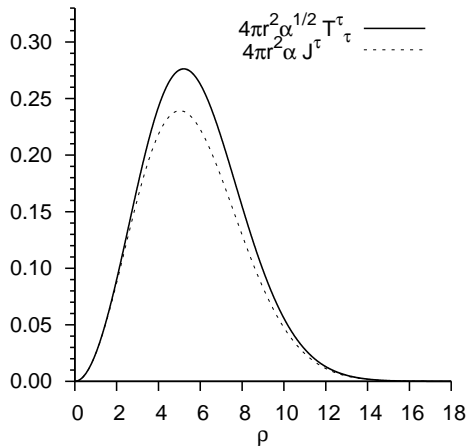}
 }
\caption{\footnotesize \label{ic}   {The matter and gravity-matter energy
    density distributions associated with a shell of radius $\rho$ on the
    initial data surface $\Sigma_0$.}}
\end{figure}

To demonstrate the significance of the freedom in specifying $\beta$,
we shall compare the result yielded in two cases, $\beta\equiv1$ and the
dynamical lapse determined by (\ref{beta_tau.eq}). The initial state is the
same in these two cases, as shown on Fig.\,\ref{ic}
\begin{figure}[h]
\begin{center}
\epsfig{figure=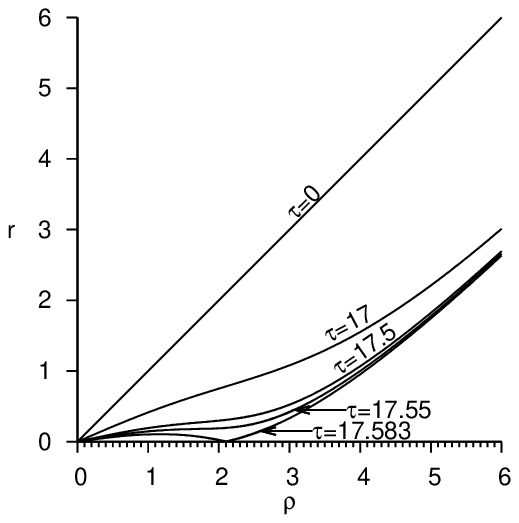,width=5.8cm}\qquad
\epsfig{figure=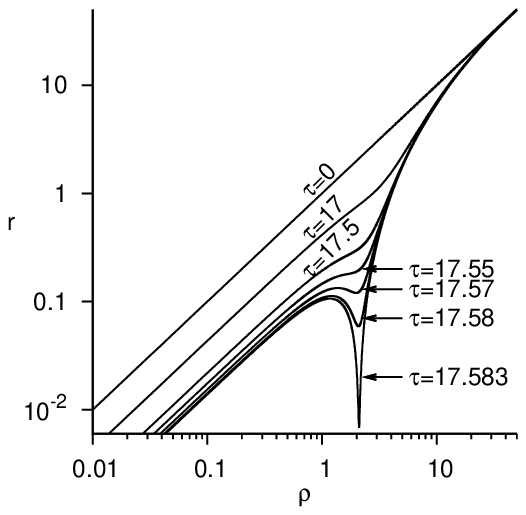,width=5.8cm}
\end{center}
\caption{\footnotesize \label{r1} The area radial coordinate function
$r(\rho)$ is plotted on various time slices linearly (left) and
logarithmically (right). A new origin can be seen to develop at $\rho_*\approx
2.1$ on both plots.}
\end{figure}
where the initial matter and gravity-matter energy density distributions
associated with a shell of radius $\rho$, $\mathscr{E}_{_{M}}\,=\,4\pi\,
r^2\sqrt{\gotR}\,{T^\tau}_\tau$ and $\mathscr{E}_{_{GM}}\,=\,4\pi\,
r^2\gotR\,\beta\,{J^\tau}$ are plotted.
The $\rho$ integrals of these quantities are, respectively, the
matter and gravity-matter energy contained in an origin-centred ball of
radius $\rho$ on the $\tau=\mathrm{const.}$ time slices,
$E_{_{M}}(\tau,0,\rho) = \int^{\rho}_{0}{\mathscr{E}_{_{M}}}\drho$ and
$E_{_{GM}}(\tau,0,\rho) = \int^{\rho}_{0}{\mathscr{E}_{_{GM}}}\drho$.

\subsubsection{Time evolution with unit lapse}

The time evolution of this initial state, with $\beta\equiv 1$, yields the
gravitational collapse of the scalar field until a spacetime singularity is
formed at $\tau_*\approx 17.583$. The change of the metric function $r$ during
the pertinent time evolution is depicted on Fig.\,\ref{r1}, where the
appearance of the second origin is clearly visible at $\rho_*\approx 2.1$. 

\begin{figure}[ht]
\unitlength1cm
 \centerline{
  \epsfysize=5.cm 
  \epsfbox{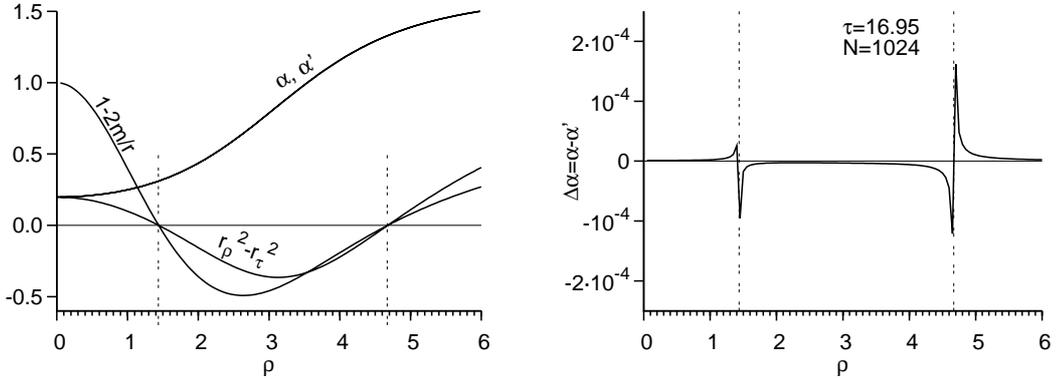}
 }
\caption{\footnotesize \label{coll1} {Functions $r_\rho^2-r_\tau^2$, $1-2m/r$,
    $\gotR$ and $\gotR'$ are plotted on a $\Sigma_\tau$ hypersurface.  On the
    right, the corresponding difference $\Delta\gotR= \gotR - \gotR'$ is
    shown. Thin vertical broken lines indicate the location of marginally
    trapped surfaces. Although the error of $\gotR'$ appears to be small, the
    spikes on the right plot indicate that its $\rho$-derivative has large
    errors at the marginally trapped surfaces.}}
\end{figure}

Fig.\,\ref{coll1}, 
shows $\gotR$, $\gotR'= (r_\rho^2-r_\tau^2)/(1-2m/r)$, $1-2m/r$ and
$r_\rho^2-r_\tau^2$ on an intermediate time level surface where trapped
surfaces are already present.  The value of $\gotR$ is calculated two ways: by
solving the evolution equation (\ref{Omega_tautau.2.eq}) and by using the
algebraic relation (\ref{Om}).  At the first sight, the two calculations yield
approximately the same result.  However, there is a tiny difference close to
the marginally trapped surfaces which, in virtue of (\ref{thpm}) and
(\ref{thpm2}), are associated with the zeros of $1-2m/r$ and
$r_\rho^2-r_\tau^2$ in the present case. This difference is enlarged by
$\rho$-derivation, thereby $\alpha_\rho$ and $\alpha'_\rho$ differ
significantly here. Whenever we started to evolve the system using
(\ref{Om}), the error became enormous in a couple of time steps, making it
impossible to carry on with the time evolution in the trapped region.  To
reach long-term stability even in regions with trapped surfaces, it turned out
to be essential to apply (\ref{Omega_tautau.2.eq}) as an evolution equation.

\begin{figure}[h]
\unitlength1cm
 \centerline{
  \epsfysize=6.4cm 
  \epsfbox{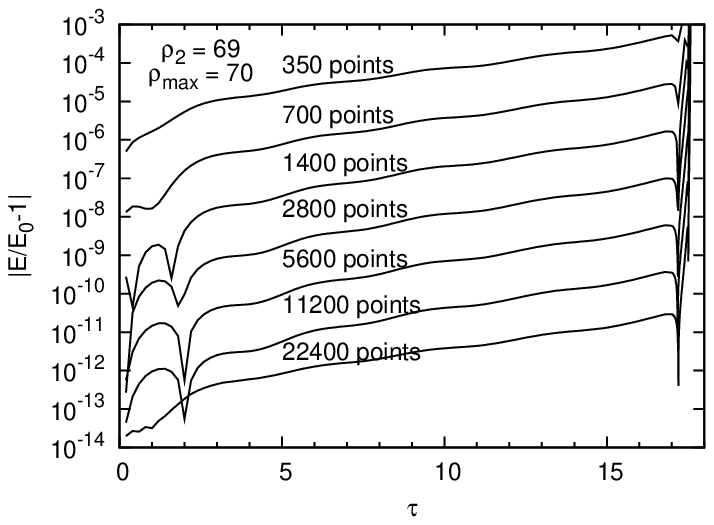}
 }
\caption{\footnotesize \label{ec}   {The validity of the energy balance
    relation (\ref{stokes2}) is shown for various spatial resolution during
    the gravitational collapse of a massive scalar field.}}
\end{figure}

Fig.\,\ref{ec} is to demonstrate that the combined gravity-matter energy is
preserved with high precision during the entire history. More
precisely, 
the energy balance relation (\ref{stokes2}), defined by making use of the
Kodama vector field in the previous section, is found to hold during the
entire evolution. The wiggly ends of the curves at $\tau\approx\tau_*$ are
associated with the very appearance of the scalar curvature singularity at
$(\rho_*,\tau_*)$. In the evaluation of the energy balance relation
(\ref{stokes2}), the particular choices $\rho_1=0$ and $\rho_2=69$ have been
made. Moreover, $E$ and $E_0$ are defined as
\begin{equation}
E\ =\ E(\tau;\,\rho_2,\,\rho_1)+S(\tau,\,0;\,\rho_2),\qquad
E_0\ =\ E(0;\,\rho_2,\,\rho_1).
\end{equation}
Notice that doubling the number of grid points shifts the curve downward by
about factor $1/16$. Therefore Fig.~\ref{ec} provides a justification that our
numerical method is fourth order accurate even with respect to $E$, which is a
complicated function of the elementary field variables. Note that the
experienced convergence order is in accordance with the applied fourth order
finite difference schemes, both for the spatial derivatives and for the time
integration.

\begin{figure}[h]
\begin{center}
\epsfig{figure=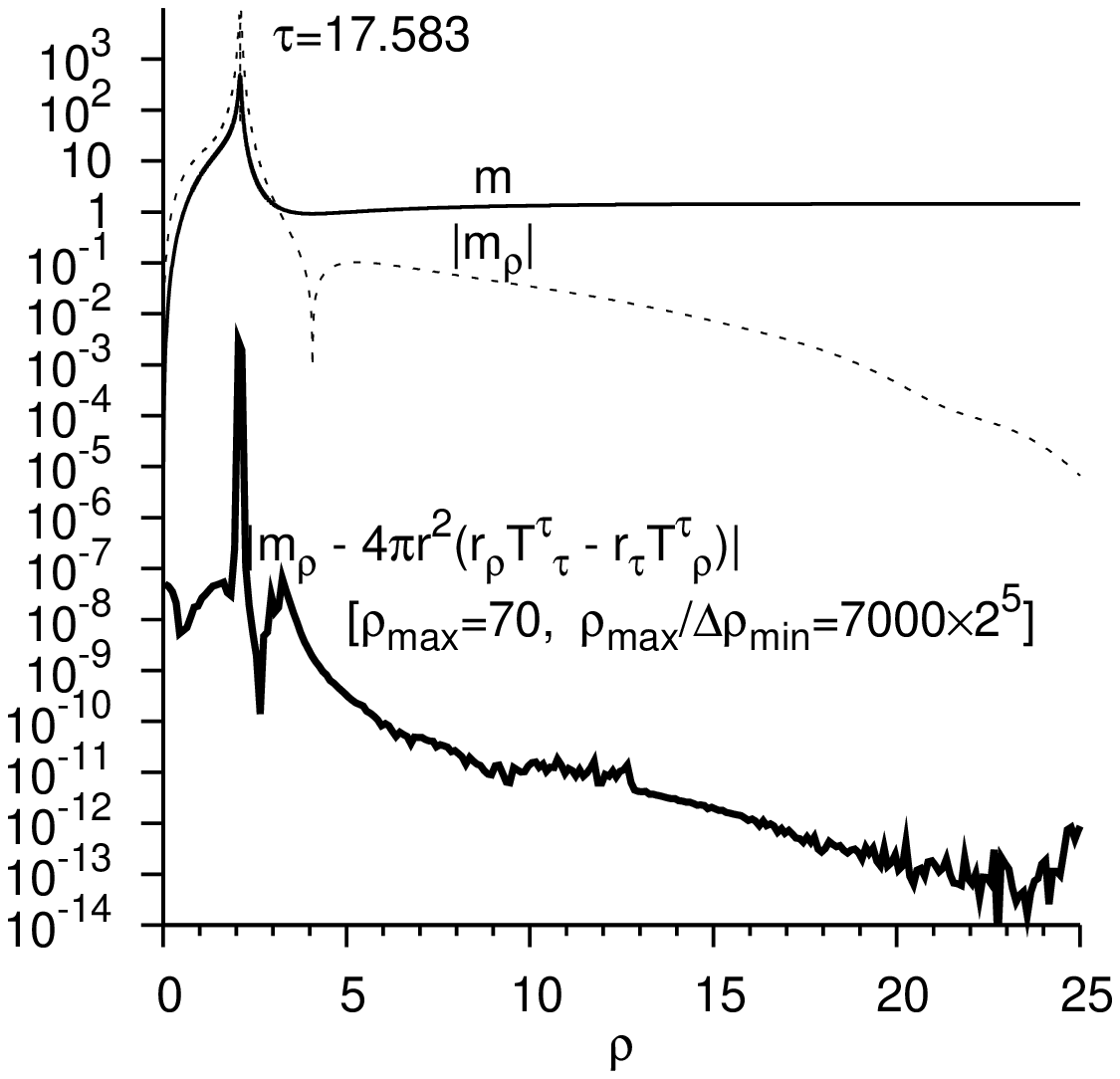,width=5.76cm}\qquad
\epsfig{figure=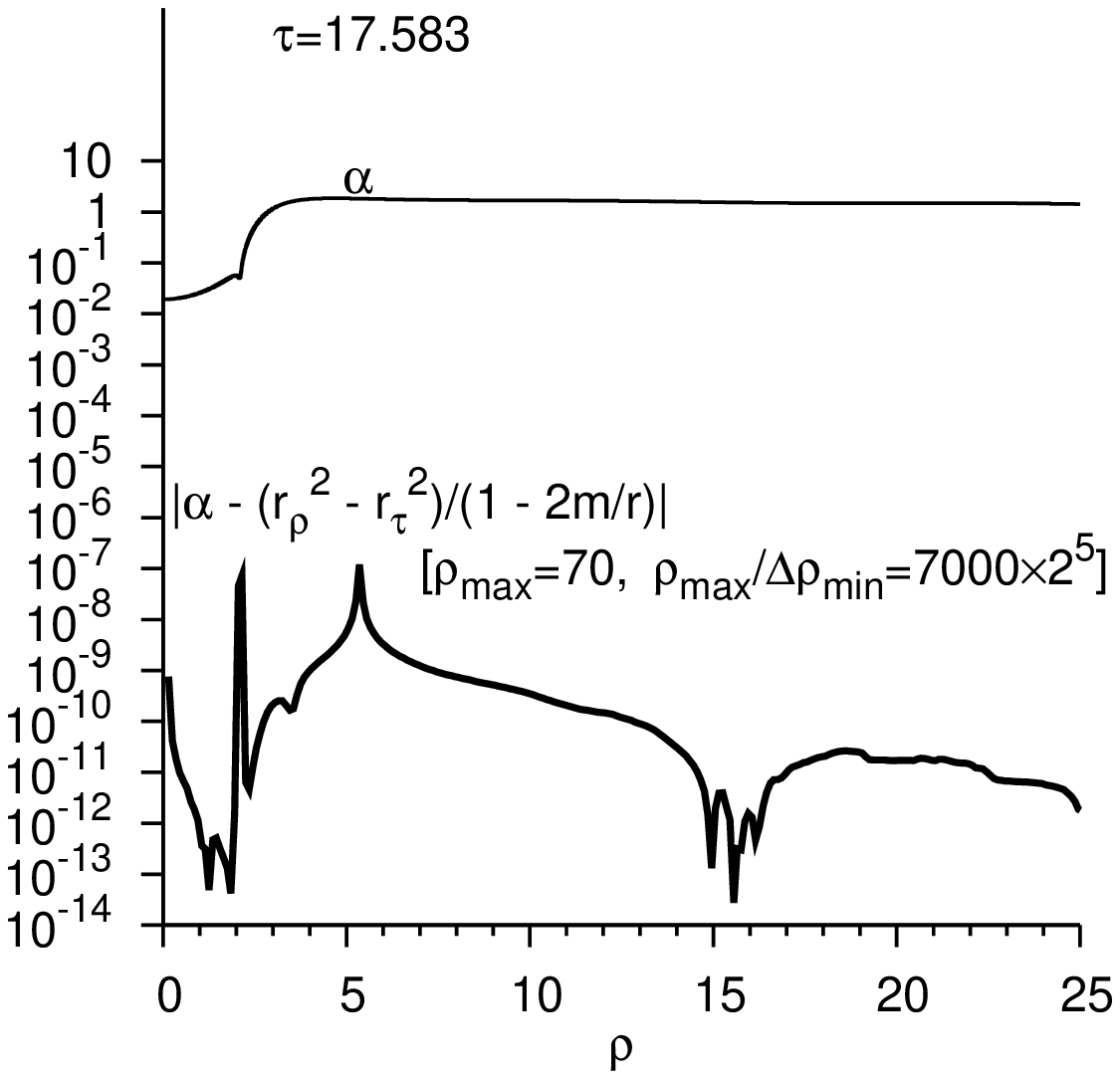,width=5.76cm}
\end{center}
\caption{\footnotesize Constraint preservation on a time level surface near
the singularity. Left: Misner-Sharp mass (narrow line), the absolute value of
its $\rho$-derivative (dotted line) and the numerical violation of constraint
(\ref{m_rho.2.eq}) (thick solid line).  Right: the $\alpha$ metric component
and the numerical violation of constraint
(\ref{Om}).}\label{constraints.beta1.fig}
\end{figure}
Fig.~\ref{constraints.beta1.fig} shows that the constraints (\ref{m_rho.2.eq})
and (\ref{Om}) are preserved with high numerical accuracy even on the last
time slice before reaching the singularity.\footnote{Note that instead of the
  $\rho$-dependent values of the constraint violations, the $\rho$-dependence
  of their averaged values are shown, in order to smooth out the noise, on
  Figs.~\ref{constraints.beta1.fig}-\ref{constraints.dynbeta.fig}. The
  averaging for each $\rho$ is done by using the interval
  $[\rho-0.05,\rho+0.05]$ around $\rho$.} The Misner-Sharp mass $m$, its
derivative, $m_\rho$, and the $\alpha$ metric component are also plotted for
comparison. Note that $m$ appears to blow up at the point where the
singularity is about to form, its maximum value reaches $m_\mathrm{max}\sim
5\times 10^2$ on the time slice shown. The error of the constraints (thick
solid curves) also have peaks in this point.

\subsubsection{Time evolution with dynamical lapse}

\begin{figure}[h]
\begin{center}
\epsfig{figure=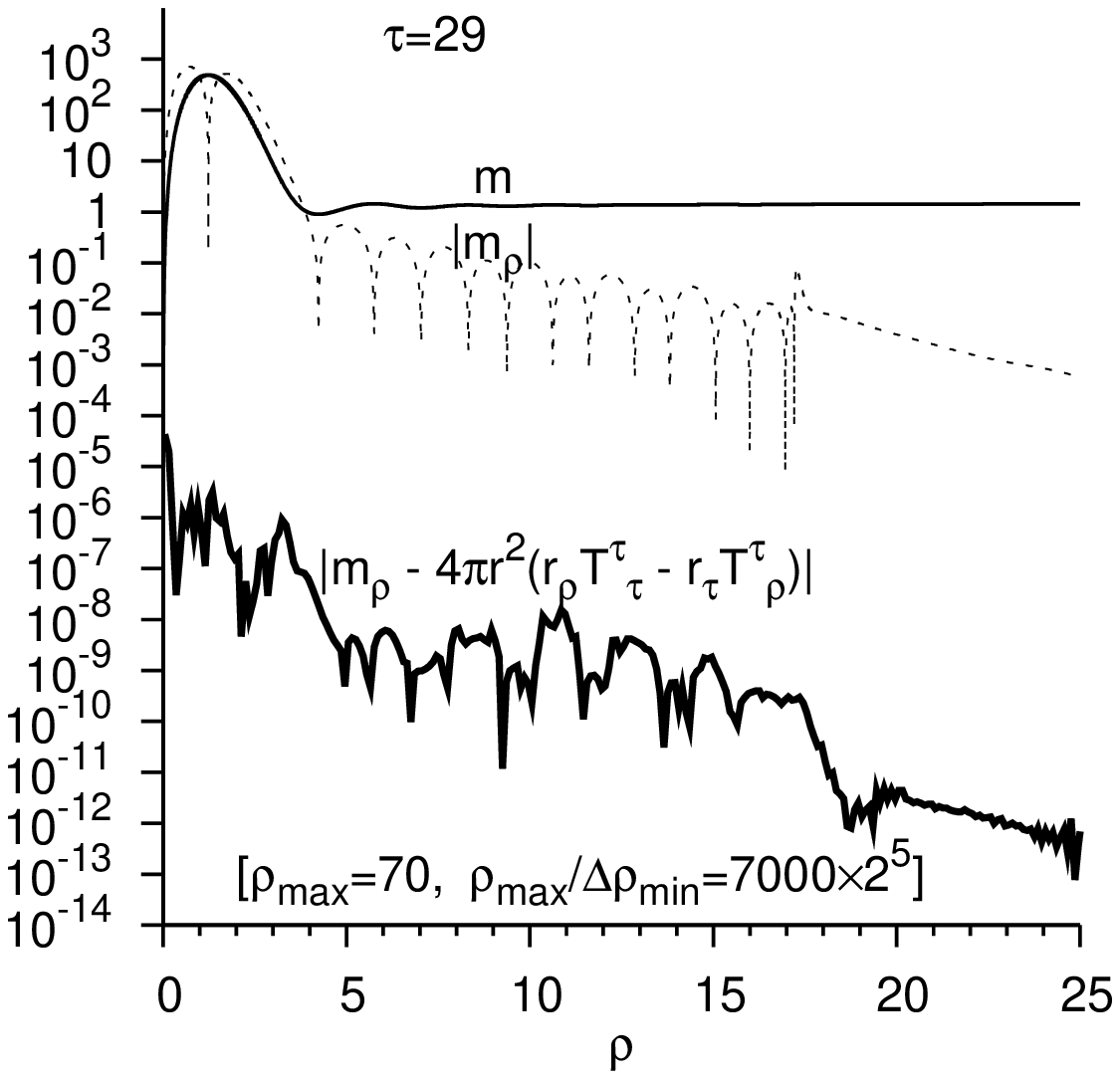,width=5.76cm}\qquad
\epsfig{figure=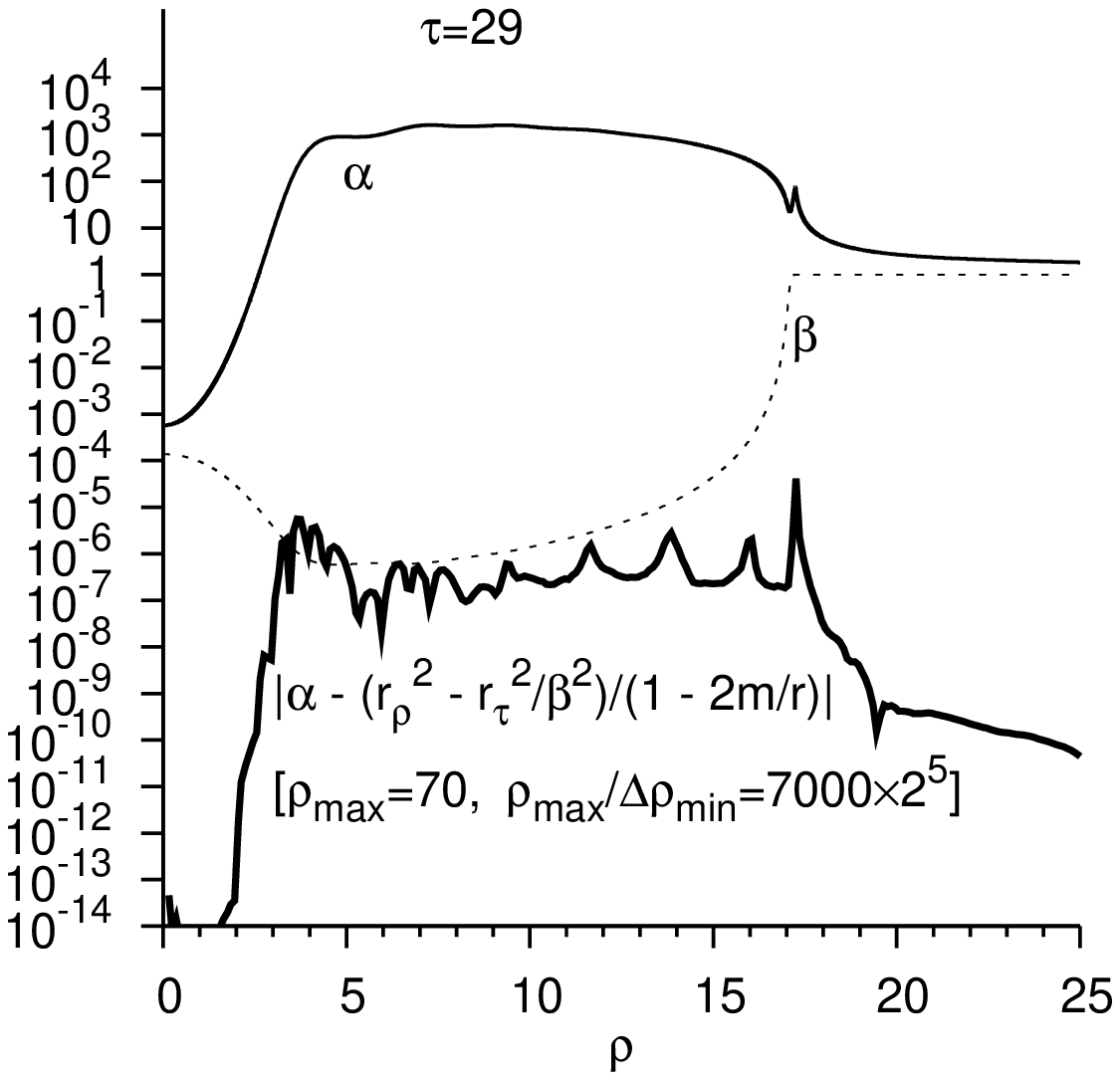,width=5.76cm}
\end{center}
\caption{\footnotesize Constraint preservation in case of dynamical lapse.}
\label{constraints.dynbeta.fig}
\end{figure}
In case of dynamical lapse, the value of the Ricci scalar is of the order
$10^{10}$ within the interval $3\lessapprox\rho\lessapprox 17$ at 
$\tau= 29$. Constraint preservation on this time level surface is shown
by Fig.\,\ref{constraints.dynbeta.fig}. 

\begin{figure}[ht]
\unitlength1cm
\centerline{
\epsfig{figure=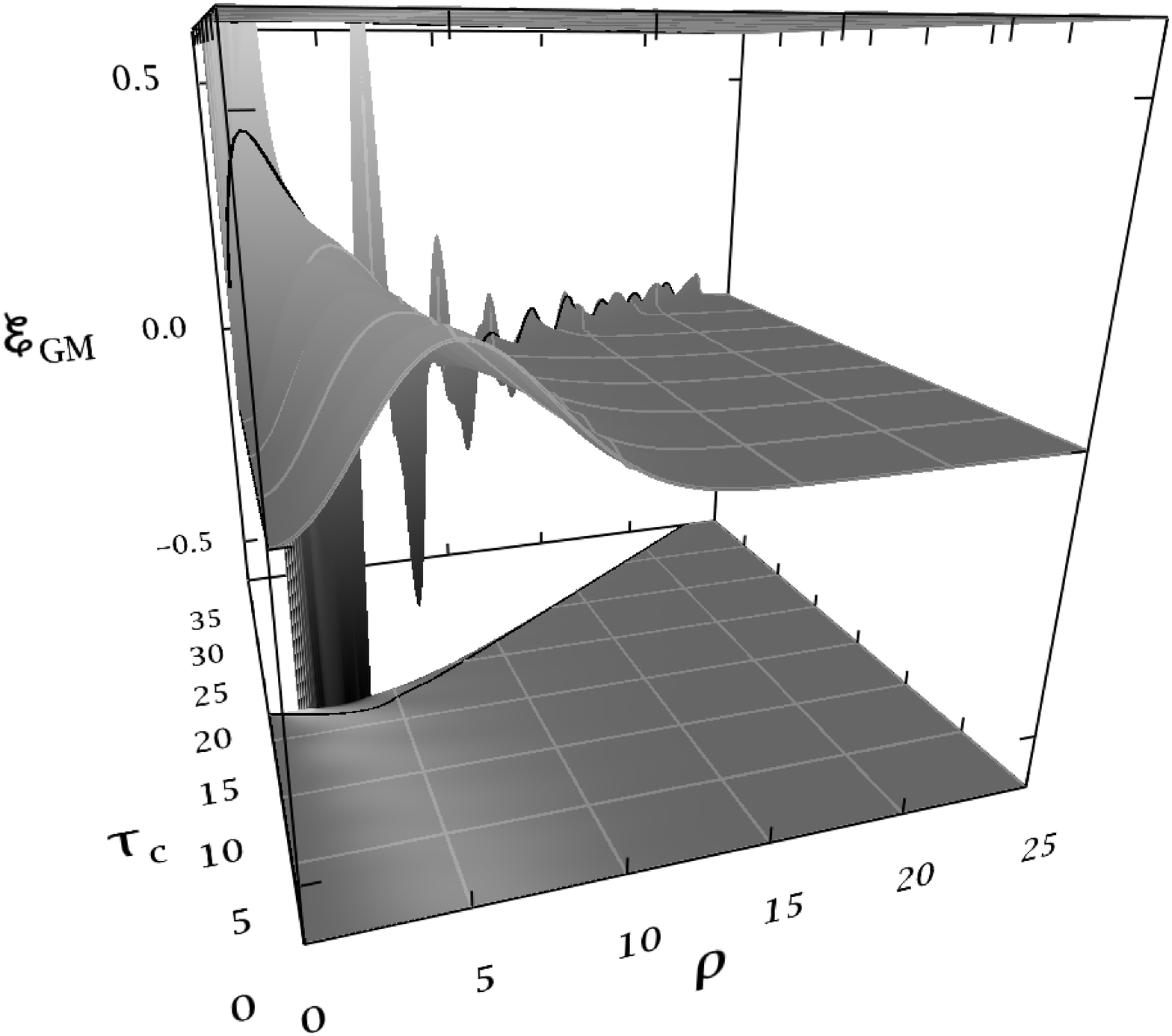,height=7cm}
 }
\caption{\footnotesize \label{ec2}    {Time evolution of the gravity-matter
    energy density distribution, $\mathscr{E}_{_{GM}}$. The black curve
    indicates the apparent horizon. {[Coloured and interactive
        versions of this and some other 3-D figures can be find
        at http://www.kfki.hu/$\sim$cspeter/numrel/2009-ekg/index.html.]} }}
\end{figure}

\begin{figure}[ht]
\begin{center}
\epsfig{figure=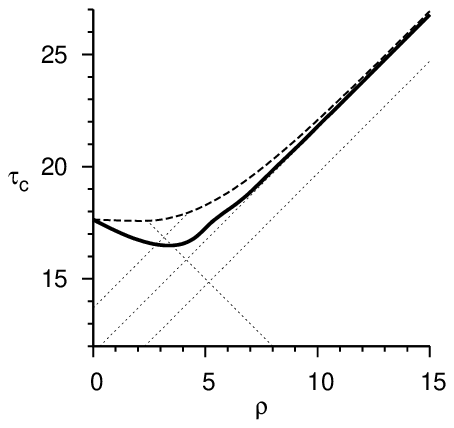,width=5.6cm}\qquad
\epsfig{figure=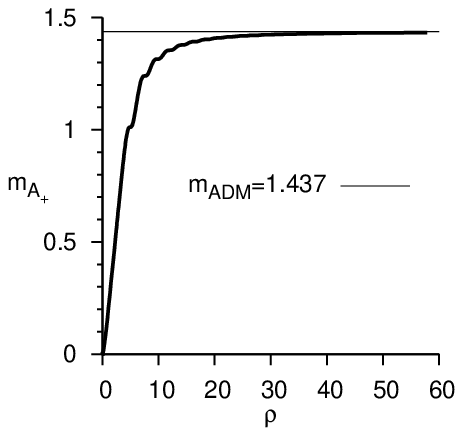,width=5.6cm}
\end{center}
\caption{\footnotesize \label{tr1} On the left, the $r=0$ singularity (dashed
 line) and the apparent horizon (continuous line) are shown.
  On the right, the $\rho$-dependence of the Misner-Sharp
  mass along this apparent horizon is plotted. }
\end{figure}

On Fig.\,\ref{ec2}  
the time evolution of the gravity-matter energy density distribution
associated with a shell of radius $\rho$, $\mathscr{E}_{_{GM}}$, is shown.
This spacetime diagram depicts the variation of $\mathscr{E}_{_{GM}}$ above
the $\rho-\tau_c$ coordinate plane, where $\tau_c$ denotes the conformal time
defined, along the constant $\rho$-coordinate lines, as $\tau_c =
\intop_0^\tau \beta \dtau$. The use of $\tau_c$ is advantageous since, as it
is indicated by Fig.\,\ref{tr1}, the radial null geodesics (see the thin
dotted lines) are well approximated\footnote{To see this recall that along
  radial null geodesics $\beta\dtau\pm \drho=0$. This, along with
  $\dtau_c=\beta\dtau+\left[\intop_0^\tau \beta_\rho \dtau\right]\drho$ and
  that the $\tau$-integral of $\beta_\rho$ is negligible, implies that in the
  $\rho-\tau_c$ coordinate plane they satisfy the differential equation
  $\dtau_c/\drho=\mp 1+\left[\intop_0^\tau \beta_\rho \dtau\right]\approx \mp
  1$.}  by $\tau_c\pm\rho= const$ lines.  On top and bottom, the
orthogonal projections of the world sheet of $\mathscr{E}_{GM}$ can be
seen. The curvy boundary represents the singularity. It is remarkable that
$\mathscr{E}_{GM}$ is positive everywhere outside the apparent
horizon. However, beyond this boundary, it starts to oscillate such that the
amplitude of this oscillation is increasing towards the intersection of the
singularity and the $\rho=0$ line. It is important to note that in spite of
this oscillatory behaviour of $\mathscr{E}_{GM}$, the Misner-Sharp mass is
always nonnegative.

On Fig.\,\ref{tr1}, the location of the apparent horizon is
plotted on the $\rho$-$\tau_c$ coordinate plane. 
In accordance with the generic expectations which are also justified by
Figs.\,\ref{coll}\,and\,\ref{coll1}, there are two marginally trapped surfaces
on each $\tau=const$ time level surface.  The $\rho$-coordinate of the inner
one is decreasing while that of the outer one is increasing.  The plotted
radial null geodesics indicate that the apparent horizon on the left panel is
everywhere achronal as it is expected.  It is also remarkable that the outer
part of the apparent horizon tends to become almost null very rapidly. On the
right panel of Fig.\,\ref{tr1} the Misner-Sharp mass is plotted as a function
of $\rho$ along the future apparent horizon, $\mycal{A}_+$. As  it is clearly
visible, $m$ is increasing monotonously along $\mycal{A}_+$ and it tends to
the value of the ADM mass $m_{_{ADM}}=m(\tau=0,\rho\rightarrow \infty)$ which
is equal to the Misner-Sharp mass $m(\tau=0,\rho_{max})$ in the present
case since the matter field is of compact support. As the area of the apparent
horizon is $A=4\pi r^2$ and $r/2 =m_{\mathscr{A}_+}\leq m_{_{ADM}}$, this plot
is also in accordance with the Penrose inequality $\sqrt{A/16\pi}\leq
m_{_{ADM}}$. 

\medskip

\begin{figure}[ht]
\begin{center}
\epsfig{figure=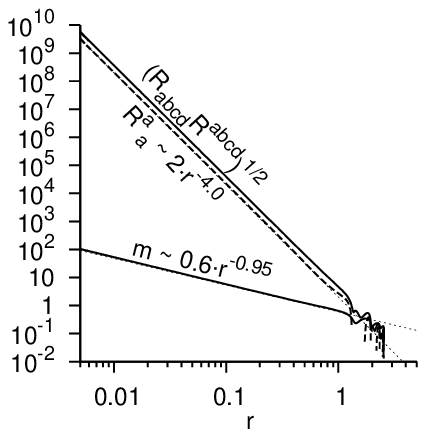,height=5.6cm}\qquad
\epsfig{figure=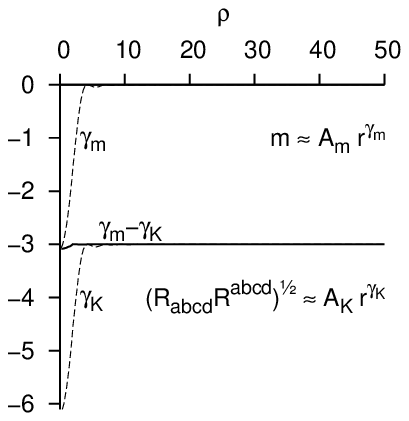,height=5.6cm}
\end{center}

\caption{\footnotesize Left: Blow up rates of the square root of the
Kretschmann scalar, $R_{abcd}\,R^{abcd}$, the Ricci scalar curvature
$R_{_{sc}}=g^{ab}R_{ab}$, and the Misner-Sharp mass $m$ along a timelike
curve, the $\rho=2.5$ line.  Right: $\rho$-dependence of the critical
exponents of the blow-up.}
\label{R3}
\end{figure}

As mentioned above, a true scalar curvature singularity develops by the end of
the time evolution inside the black hole region when both the Kretschmann
scalar, $R_{abcd}\,R^{abcd}$, and the Ricci scalar curvature,
$R_{_{sc}}=g^{ab}R_{ab}$, tend to infinity.  As it is justified by the log-log
plots on left panel of Fig.\,\ref{R3}, the blow up rate of
$R_{abcd}\,R^{abcd}$, $R_{_{sc}}=g^{ab}R_{ab}$ and $m$, at fixed $\rho=2.5$
location, are $r^{-8}$, $r^{-4}$  and $r^{-0.95}$, respectively. On the right
panel of Fig.~\ref{R3}, the $\rho$-dependence of the critical exponents
$\gamma_m$, $\gamma_K$ and $\gamma_K-\gamma_m$ of the blow up rates $m\approx
A_m\,r^{\gamma_m}$ and $(R_{abcd}\,R^{abcd})^{1/2}\approx A_K\,r^{\gamma_K}$
are shown. It is visible that these curvature scalars, along with the
Misner-sharp mass, diverge faster close to  the origin.  We would like to
point to the fact that the blow-up of the Kretschmann scalar is completely
consistent with the relation $({R_{abcd}\,R^{abcd}})^{1/2}\geq
4\sqrt{2}\,m/r^3$ derived by Christodoulou for the case of the collapse of a
massless scalar field in \cite{christ1}.

Interestingly, in spite of the divergence of the Ricci scalar, the quantity
$\gotR\beta\,r^2R_{_{sc}}$ remains finite. Thereby the Einstein-Hilbert action  
\begin{equation}\label{E-H}
{S}_{_{EH}}=\int_{M'=\Sigma'\times [\tau_0,\tau_*)}
  R_{_{sc}}\,\epsilon=4\pi\int_{\tau_0}^{\tau_*}\int_{\rho_1}^{\rho_2}
  \gotR\beta\,r^2R_{_{sc}} 
  \,\drho\,\dtau\,, 
\end{equation}
remains finite as well, for any particular choice of $\rho_1, \rho_2$ with
$0\leq\rho_1<\rho_2<\rho_{max}$ in
$\Sigma'=[\rho_1,\rho_2]\times\mathbb{S}^2$. 

\subsection{Gravitational collapse with spatial topology
  $\Sigma_0=\mathbb{S}^3$}\label{S3coll}

Whenever the topology of the initial data surface, $\Sigma_0$, is chosen to be
$\mathbb{S}^3=[0,\pi]\times\mathbb{S}^2$, there has to be two origins on
$\Sigma_0$ in the considered spherically symmetric setting.  Motivated by and
mimicking the simplicity of the geometric setup of the FRW cosmological model,
we used procedure B (see section \ref{inidata}) and chose the
functional form of the initial $\gotR$ and $r$ as
\begin{equation}
\gotR\,=\,R_0^2\quad {\rm and}\quad r\,=\,R_0\sin\rho,
\label{homogen_psi_wave0.initial.eq}
\end{equation}
To have a nearly central symmetric time evolution, the initial data for the
scalar field was chosen to be slightly asymmetric by requiring $\psi$ and
$\psi_\tau$ to take the form  
\begin{equation}
\psi\,=\,\psi_0\cos n\rho + \psi_1\cos n'\rho \quad {\rm and}\quad \psi_\tau\ 
=\,\dot\psi_0\cos n\rho + \dot\psi_1\cos n'\rho\,,
\label{homogen_psi_wave1.initial.eq}
\end{equation}
\begin{figure}[ht]
\begin{center}
\epsfig{figure=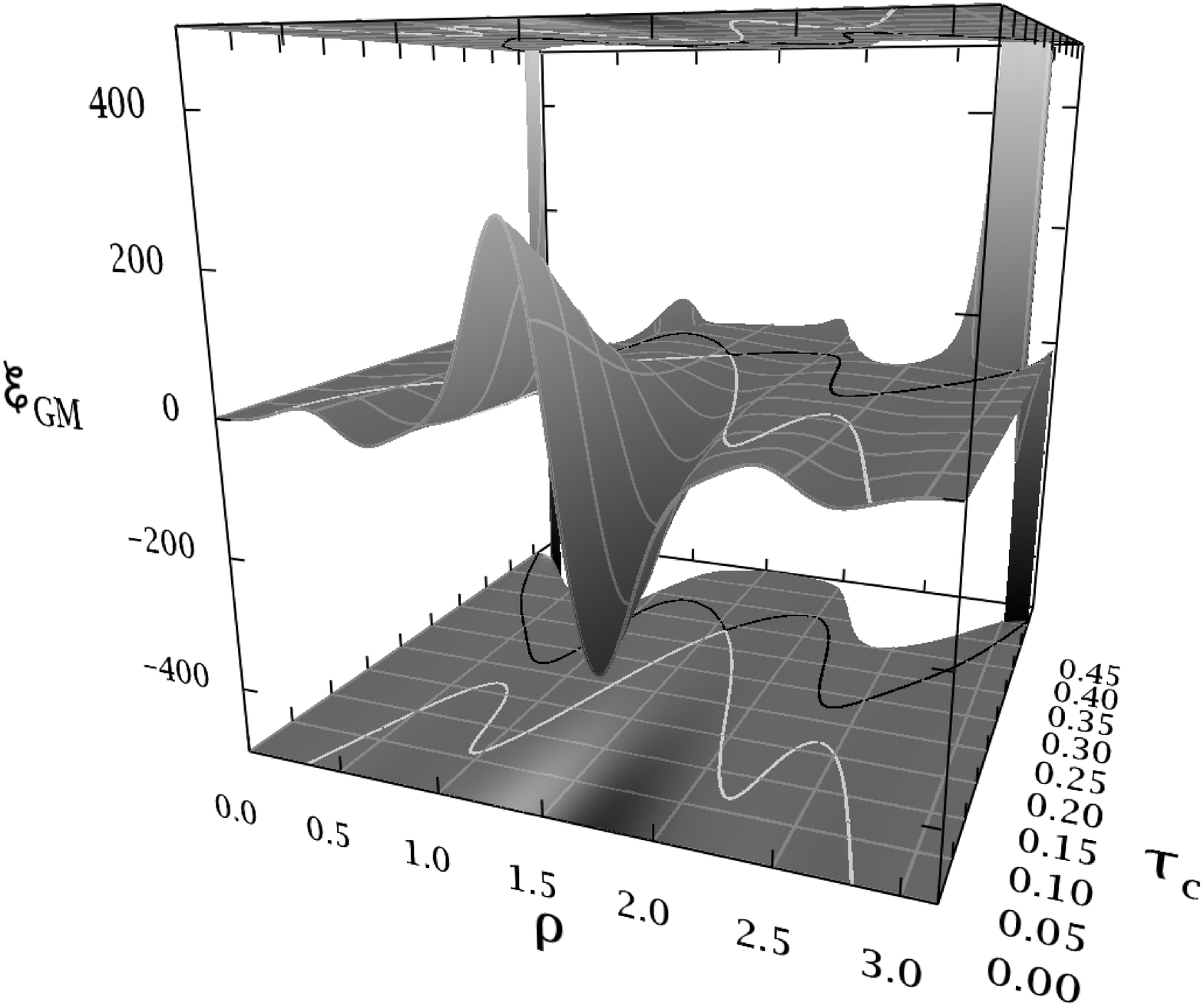,height=7.cm}
\end{center}
\caption{\footnotesize \label{S3-energy} The time evolution of
  $\mathscr{E}_{_{GM}}$. The black and white curves indicate the location of
  the future and past apparent horizons.}
\end{figure}
The parameters $R_0$, $\psi_0$, $\psi_1$, $\dot\psi_0$, $\dot\psi_1$, $n$ and
$n'$ in the above functional expressions were fixed as 
\begin{equation}\label{S3par}
R_0=10,\ \psi_0=0.4,\ \psi_1=0.005,\ \dot\psi_0=1.2,\ \dot\psi_1=0.01,\ n=3
\ \ {\rm and}\ \ n'=2\,, 
\end{equation}
while the number of the spatial grid points $N$ was chosen to be
1,000.

On Fig.\,\ref{S3-energy}
the time evolution of the gravity-matter energy density distribution
associated with a shell of radius $\rho$, $\mathscr{E}_{_{GM}}$, is shown.  
\begin{figure}[ht]
\begin{center}
\epsfig{figure=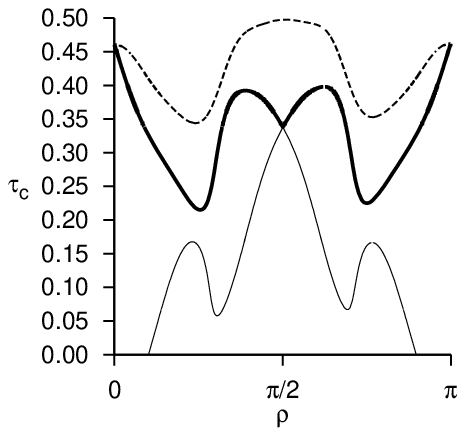,height=5.8cm}\qquad
\epsfig{figure=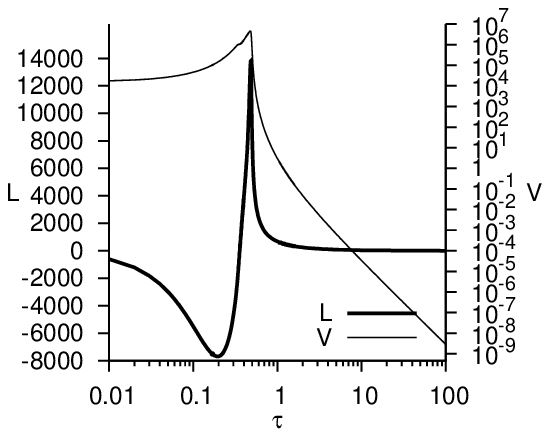,width=6.8cm}
\end{center}
\caption{\footnotesize \label{S3-mt-action} Left: The location of the future
  (thick solid line) and the past (thin solid line) apparent horizons, along
  with the singularity (dashed line). Right: The time dependence of the
  $3$-volume, $V=\int_0^\pi\sqrt{\alpha}\beta\drho$, and the Lagrangian,
  $L=4\pi\int_0^\pi\alpha\beta r^2 R_{_{sc}}\drho$.}
\end{figure}
The left panel of Fig.\,\ref{S3-mt-action} shows a somewhat exotic
distribution of the marginally trapped surfaces produced
by the time evolution. This plot justifies that the formation of the curvature
singularity is censored, i.e., a connected future trapped region develops in
advance to the appearance of the singularity.
There are two apparently smooth curves starting at the bottom and
ending at the opposite upper corner where $\theta_-$ and $\theta_+$ are
identically zero, respectively. These curves intersect at $\rho=\pi/2$. The
corresponding $2$-sphere is a {maximal} surface. The points on the thick part of
the curves represent future marginally trapped $2$-surfaces while the thin
parts depict the location of the past marginally trapped
$2$-surfaces. Thereby the associated past apparent horizon may also be
referred as the boundary of the dynamical white hole region.  
  
On the right panel of Fig.\,\ref{S3-mt-action} the time dependence of the
spatial $3$-volume, $V=\int_0^\pi\sqrt{\alpha}\beta\drho$, is shown.  It
increases exponentially up to $\tau\approx 0.48$ after which the tendency is
reversed and an even faster decay of $V$ is experienced. 

\begin{figure}[ht]
\begin{center}
\epsfig{figure=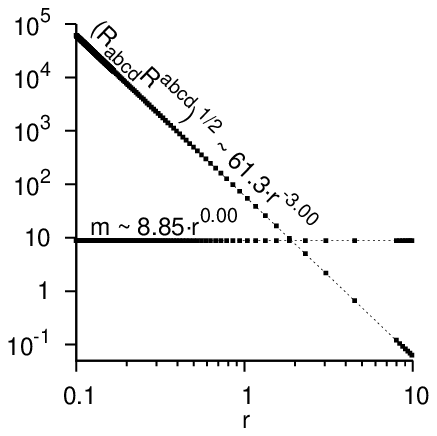,height=5.5cm}\qquad
\epsfig{figure=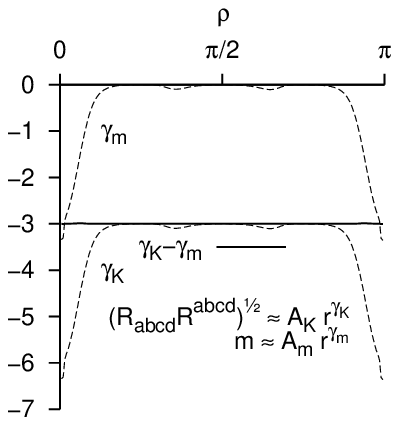,width=5.8cm}
\end{center}
\caption{\footnotesize \label{S3baby-r} Left: The $r$-dependence of
  $(R_{abcd}\,R^{abcd})^{1/2}$ and the Misner-Sharp mass $m$ along a timelike
  curve, the $\rho=2.5$ line.  Right: $\rho$-dependence of the critical
  exponents.}
\end{figure}

An interesting new feature of the evolution is that the metric function
$\gotR$ appears to blow up while approaching the singularity.
However, its blow-up is compensated by the decay of $\beta$;
the proper time $t(\rho)=\int_0^{\tau_*}\sqrt{\gotR}\beta\,\dtau$ was
found to be finite along any constant $\rho$ world-line---meaning that the
singularity is in finite proper time distance from the initial data surface.

The $r$-dependence of the square root of the Kretschmann scalar and the
Misner-Sharp mass, along the world-line $\rho=2.5$, are indicated on the left
panel of Fig.\,\ref{S3baby-r}. On the right the $\rho$-dependence of the
critical exponents $\gamma_K$, $\gamma_m$ and $\gamma_K-\gamma_m$ are
given. Interestingly, Christodoulou's relation
$({R_{abcd}\,R^{abcd}})^{1/2}\geq 4\sqrt{2}\,m/r^3$---derived only  for the
case of the collapse of a massless scalar field in a spacetime with spatial
topology $\mathbb{R}^3$ in \cite{christ1}---does apply here as well.

The value of the Einstein-Hilbert action,
\begin{equation}\label{E-H2}
{S}_{_{EH}}=\int_{M'={\mathbb{S}^3}\times [0,\tau_*)}
  R_{_{sc}}\,\epsilon=\int_{0}^{\tau_*} L\,\dtau\,,
\end{equation}
remains finite as it is indicated by the $\tau$-dependence of the Lagrangian
$L=4\pi\int_{0}^{\pi} \gotR\beta\,r^2R_{_{sc}} \,\drho$, shown on the
left panel of Fig.\,\ref{S3-mt-action}.

\subsection{Gravitational collapse with spatial topology
  $\Sigma_0=\mathbb{S}^1\times\mathbb{S}^2$}\label{S1xS2coll}

Whenever the topology of the initial data surface $\Sigma_0$ is
$\mathbb{S}^1\times\mathbb{S}^2$, the simplest initial configuration to start
with is a homogeneous `torus' with constant initial radius and with constant
initial expansion. 
Accordingly, we used scheme A and chose the initial $r$ and $r_\tau$ as
\begin{equation}
r(\rho)\,=\,r_0 \quad {\rm and} \quad r_\tau(\rho)\,=\,\dot r_0\,,
\end{equation}
while the number of the spatial grid points $N$ was chosen to be 4,000.
\begin{figure}[ht]
\begin{center}
\epsfig{figure=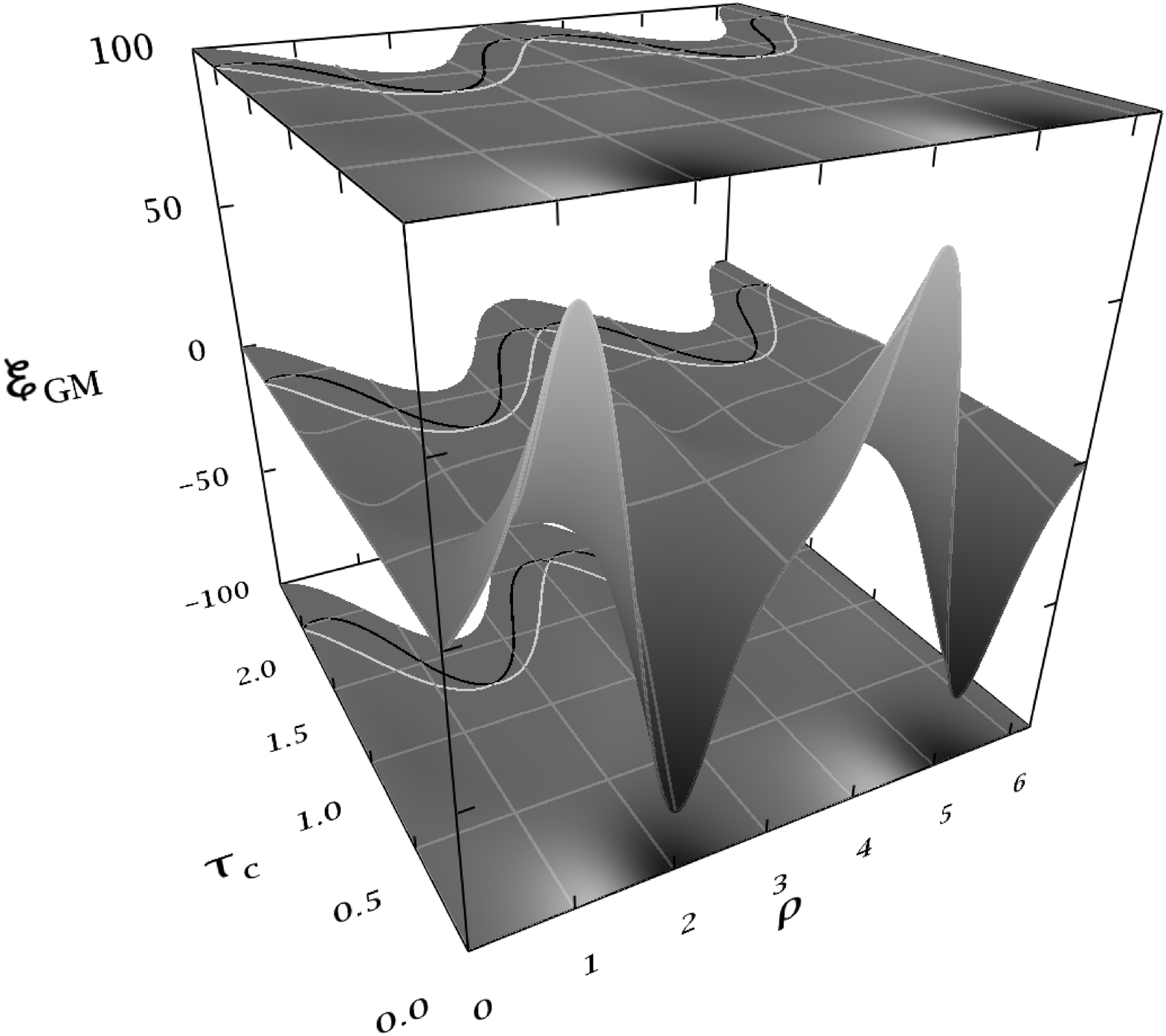,height=7.cm}
\end{center}
\caption{\footnotesize \label{S1xS2-energy} The time evolution of
  $\mathscr{E}_{_{GM}}$. The black and white curves indicate the location of
  the future and past apparent horizons.}
\end{figure}

For the the initial $\psi$ and $\psi_\tau$, we used the relations in
(\ref{homogen_psi_wave1.initial.eq}) with the parameter values in
(\ref{S3par}), while the values of the metric parameters were chosen as
\begin{equation}
r_0=1,\quad{\rm and}\quad \dot r_0=1\,.
\end{equation}

It is important to keep in mind that now $\rho$ is a periodic coordinate,
along the $\mathbb{S}^1$ factor. This periodicity length was chosen to be
$2\pi$. Notice also that according to the above choice of initial data, there
is  no origin at all on $\Sigma_0$.

\begin{figure}[ht]
\begin{center}
\epsfig{figure=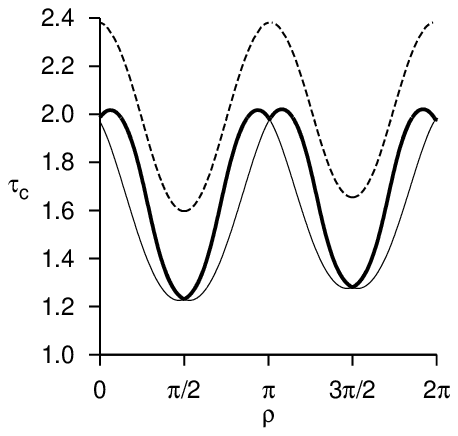,width=6.cm}\qquad
\epsfig{figure=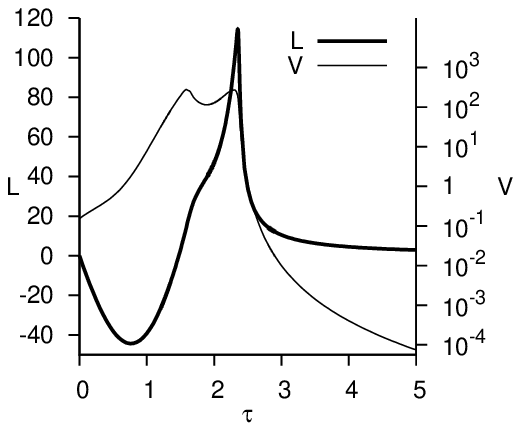,width=6.7cm}
\end{center}
\caption{\footnotesize \label{S1xS2-mt-action} Left: The $\rho-\tau_c$
  coordinate plane with the singularity (dashed curve), and the future (thick
  curve) and past (thin curve) apparent horizons.  Right: The
  $\tau$-dependence of the $3$-volume,
  $V=\int_0^{2\pi} \sqrt{\alpha}\beta\drho$, and the Lagrangian,
  $L=4\pi\int_0^{2\pi}\alpha\beta r^2 R_{_{sc}}\drho$.}
\end{figure}

On Fig.\,\ref{S1xS2-energy} the time evolution of the gravity-matter energy
density distribution associated with a shell of radius $\rho$,
$\mathscr{E}_{_{GM}}$, is shown. As opposed to the other cases,
$\mathscr{E}_{_{GM}}$ tends to zero at the singularity. The Kretschmann scalar
blows up in the singularity such that Christodoulou's relation
$({R_{abcd}\,R^{abcd}})^{1/2} \geq 4\sqrt{2}\,m/r^3$ holds.
The location of the future and past apparent horizons are also
indicated on the left panel of Fig.\,\ref{S1xS2-mt-action} .

The appearance of curvature singularities are preceeded by the formation of
trapped regions. As in the previous case,   both future and past apparent
horizons are formed. They intersect at the $2$-surfaces with $\rho\approx
0,\pi/2,\pi,3\pi/2$. {Interestingly enough, while the $2$-surface with
  $\rho\approx0$ and $\rho\approx\pi$ are maximal the other two $2$-surfaces
  have no definite  character. More precisely, they are locally maximal in the
  spatial $\rho$-direction whereas they are locally minimal in the timelike
  $\tau$-direction.} The spacetime region with untrapped surfaces is very
limited in the present case. It is represented by the points between the
future (thick solid line) and the past (thin solid line) apparent horizons on
the left panel of Fig.\,\ref{S1xS2-mt-action}. 

The $\tau$-dependence of the Lagrangian $L=4\pi\int_{0}^{2\pi}
\gotR\beta\,r^2R_{_{sc}} \,\drho\,,$ indicates (see the right panel of
Fig.\,\ref{S1xS2-mt-action}) that the Einstein-Hilbert action,
\begin{equation}\label{E-H3}
{S}_{_{EH}}=\int_{M'={\mathbb{S}^1\times\mathbb{S}^2}\times [0,\tau_*)}
  R_{_{sc}}\,\epsilon=\int_{0}^{\tau_*} L\,\dtau\,,
\end{equation}
remains finite as in the previous cases.

Finally, the $\tau$-dependence of the spatial $3$-volume,
$V=\int_0^{2\pi}\sqrt{\alpha}\beta\drho$, is also shown on the right.  The
volume starts to increase rapidly, then a short oscillation followed by
an extremely fast contraction can be seen.

\section{Summary}\label{final}

In most of the former numerical simulations, one of the aims was to minimise
the extent of the trapped region. This was achieved by making use of
techniques like singularity (and trapped region) avoiding slicings and black
hole excision---the latter was originally suggested by Unruh in 1984 (for its
first adaptation in numerical simulations see, e.g., \cite{exc1,exc2}).  The
unsatisfactory aspect of this trouble avoiding attitude, and the reason for
choosing here the simplest possible framework of spherically symmetric
dynamical configurations are justified by the following comments of David
Hilbert (1902):  

\smallskip
{\it "In dealing with mathematical
  problems, specialisation plays, as I believe, a still more important part
  than generalisation. Perhaps in most cases where we unsuccessfully seek the
  answer to a question, the cause of the failure lies in the fact that
  problems simpler and easier than the one in hand have been either
  incompletely solved, or not solved at all. Everything depends then, on
  finding those easier problems and on solving them by means of devices as
  perfect as possible and of concepts capable of generalisations."}
\smallskip

The key technical achievements that, as we believe, have not been applied
before in numerical simulations, are the following: 
\begin{itemize}
\item[(1)] A strongly hyperbolic (symmetrisable) first order system
  of evolution equations was singled out for $4$-dimensional spherically
  symmetric gravitating systems. 
\item[(2)] The analytic setup ensures that time evolution can be studied on
  equal footing in trapped and untrapped regions.
\item[(3)] The numerical framework applies this analytic setup and incorporates
the techniques of AMR.
\end{itemize}
By making use of these technical developments we achieved the following
results:
\begin{itemize}
\item[(1)] By introducing a suitable evolution equation for the lapse function
$\beta$, the extent of the investigated spacetime domain was enlarged
significantly and the physical singularities were approached arbitrarily closely
everywhere.
\item[(2)] The location of the future and past
apparent horizons has been determined. 
\item[(3)] Detailed investigation of the rate
of curvature blow-up while approaching the singularity. 
\item[(4)] The Einstein-Hilbert action remained finite in all the investigated
cases, in spite of the blow up of the Ricci scalar.
\end{itemize}

\medskip

As one of our aims was to clear up some conceptual issues of topology
change, let us close this paper by related comments. 

In dealing with the problem of topology changes, one should start by recalling
the result of Geroch \cite{geroch} which asserts that if topology change
develops then there must exist either closed causal curves beyond the  Cauchy
horizon\footnote{As it was discussed in the introduction in the present
context one could always think of the limit of the time level surfaces as the
Cauchy horizon.}  or a spacetime singularity has to appear.  Due to Tipler's
theorem \cite{tipler}, the latter case manifests itself if Einstein's
equations are imposed.  Based on these results, it has been widely held that
no indication of topology changes will ever show up in classical general
relativity, therefore the quantum theory should be investigated to see whether
they may occur. 

\medskip

The idea that the topology may change in quantum gravity was originally
proposed by Wheeler \cite{wheeler,wheeler2}.  Since then, one of the most
important questions in any quantum theory of gravity is whether there is a
non-zero probability for the topology of space to change. There have been a
number of classical investigations aiming to demonstrate the feasibility of
topology change by making use of the techniques of differential geometry and
topology (see, e.g., \cite{borde, horowitz,dowkergarcia,Sorkin}).
Nevertheless, no quantitative investigations have been carried out yet.
Therefore it is important to emphasise that some of our findings provide the
first definite quantitative support to all the former speculations concerning
the existence of topology changes.

\medskip

In summarising our pertinent observations, we can say, in accordance with the
above recalled results of Geroch and Tipler, that instead of having a regular
Cauchy horizon, along with a causality violating region beyond, a spacetime
singularity develops at the ``new origins''. It was also found that the
Kretschmann scalar always blows up there. It is important to emphasise that
the Einstein-Hilbert action 
$
{S}_{_{EH}}=\int_{M=\Sigma\times [\tau_0,\tau_*)}
  R_{_{sc}}\epsilon
$
was found to be bounded in all of our investigations.

\medskip

The existence of time developments with apparent topology changes may be
significant in quantum theoretical considerations. For instance, in the sum
over histories approach to quantum gravity, the transition amplitude for the
topology change between the two Riemannian manifolds $(\Sigma_1,h_1)$ and
$(\Sigma_2,h_{2})$ is supposed to be given by the formula
\begin{equation}\label{fint}
\left<\,(\Sigma_1,h_1)\,|\,(\Sigma_2,h_{2})\,\right>=\sum_{M}\,\int_{{}^4
\mathcal{G}} \exp\left[\,i\, {S}_{_{EH}}\right] \,\mathcal{D}g_{ab}\, ,
\end{equation}
where the boundary of $M$ is the disjoint union of $\Sigma_1$ and $\Sigma_2$,
${S}_{_{EH}}$ denote the Einstein-Hilbert action and ${}^4 \mathcal{G}$ is the
space of $4$-dimensional Lorentzian geometries on $M$, while the sum is over
all $4$-manifolds whose boundary is the disjoint union of $\Sigma_1$ and
$\Sigma_2$ with Riemannian metrics $h_1$ and $h_2$ (see, e.g.,
\cite{wheeler,wheeler2,hawking,horowitz,dowker}). 

Admittedly, the right hand side of (\ref{fint}) is completely formal and is
far from being defined as yet. Nevertheless, regardless of the specific form
of the measure on the space of metrics, $\mathcal{D}g_{ab}$, it is widely held
that not only the smooth Lorentzian metrics but all the metrics with finite
Einstein-Hilbert action  should be taken into account in evaluating the
functional integral. Therefore the spacetimes with topology changes
investigated in this paper, might be of interest in quantum theoretical
considerations.

It is of obvious importance to know whether the methods introduced in this
paper could be adopted in more generic geometrical setup allowing the presence
of gravitational waves and with the inclusion of binary black holes. {We
  would like to mention that by making use of the analytic setup suggested in
  \cite{sanchez} such a generalisation seems to be possible. The results of
  our corresponding investigations will be published elsewhere.} 

\section*{Acknowledgements} 
This research was supported in part by OTKA grant K67942.


\end{document}